\documentclass[]{aastex631}
\usepackage{amsmath}

\begin{document}

\title{{Substantial} extension of the lifetime of the terrestrial biosphere}

\author[0000-0001-9289-4416]{R.J. Graham}

\affiliation{Department of Geophysical Sciences, University of Chicago\\            
5734 S. Ellis Ave.\\ 
Chicago, IL 60637, USA}
\correspondingauthor{R.J. Graham}
\email{arejaygraham@uchicago.edu}
\author[0000-0002-7325-8139]{Itay Halevy}
\affiliation{Department of Earth and Planetary Sciences, Weizmann Institute of Science \\
234 Herzl St.\\
Rehovot, Israel}

\author[0000-0001-8335-6560]{Dorian Abbot}
\affiliation{Department of Geophysical Sciences, University of Chicago\\            
5734 S. Ellis Ave.\\ 
Chicago, IL 60637, USA}



\begin{abstract}
Approximately one billion years (Gyr) in the future, as the Sun brightens, Earth's carbonate-silicate cycle is expected to drive CO$_2$ below the minimum level required by vascular land plants, eliminating most macroscopic land life. Here, we couple global-mean models of temperature- and CO$_2$-dependent plant productivity for C$_3$ and C$_4$ plants, silicate weathering, and climate to re-examine the time remaining for terrestrial plants. If weathering is weakly temperature-dependent (as recent data suggest) and/or strongly CO$_2$-dependent, we find that the interplay between climate, productivity, and weathering causes the future luminosity-driven CO$_2$ decrease to slow and temporarily reverse, averting plant CO$_2$ starvation. This dramatically lengthens plant survival from 1 Gyr up to $\sim$1.6-1.86~Gyr, until extreme temperatures halt photosynthesis, suggesting a revised kill mechanism for land plants and potential doubling of the future lifespan of Earth's land macrobiota. An increased future lifespan for the complex biosphere may imply that Earth life had to achieve a smaller number of ``hard steps'' (unlikely evolutionary transitions) to produce intelligent life than previously estimated. These results also suggest that complex photosynthetic land life on Earth and exoplanets may be able to persist until the onset of the moist greenhouse transition. 

\end{abstract}



\section{Introduction}\label{sec:intro}
In the far future, as the Sun brightens, Earth's surface will warm, and in response the carbonate-silicate cycle \citep{Walker1981} is expected to draw CO$_2$ out of the atmosphere through climate-dependent silicate weathering and carbonate burial \citep{Caldeira1992}. This will create an increasingly stressful environment for land plants, eventually driving them extinct through CO$_2$ starvation, at the CO$_2$ compensation point, or through overheating, at their upper temperature threshold \citep{Lovelock1982,Caldeira1992}. This would also lead to the extinction of macroscopic life on land that relies on land plants \citep{Judson2017}. Most previous work has concluded that CO$_2$ starvation is the more likely kill mechanism and that this is likely to occur approximately 1 billion years (Gyr) in the future (Table \ref{tab:lifespan_studies}).

The rate of silicate weathering at the planetary scale is usually assumed to vary exponentially with temperature, with an e-folding scale of $T_\mathrm{e}$, and to have a power-law dependence on CO$_2$, with an exponent $\beta$ (Eq. \ref{eqn:weathering}). Most previous work (Table \ref{tab:lifespan_studies}) has assumed that silicate weathering is strongly temperature-dependent ($T_\mathrm{e}\approx$10--20~K) and weakly CO$_2$-dependent ($\beta\approx$0.25-0.5). Recent studies instead support a much weaker effective temperature dependence, with $T_\mathrm{e}$=30--40~K \citep{Maher2014,KrissansenTotton2017,Winnick2018,Graham2020,Herbert2022,Brantley2023}. The CO$_2$ dependence of silicate weathering has received less attention than the temperature dependence, but a range of $\beta$=0.2--0.9 is allowed by available data \citep{KrissansenTotton2017,Palandri2004,Winnick2018,Hakim2021}.

Life's expansion onto land likely increased the dissolution rate of silicate minerals at Earth's surface under a given set of climate conditions, leading to lower equilibrium CO$_2$ and lower surface temperatures to maintain balance between weathering and outgassing \citep{Volk1987,Berner1992,Schwartzman2017}. This biotic weathering enhancement results mostly from plant acidification (through organic acid excretion and increased belowground respiration) and stabilization of soils, as well as changes in water cycling \citep{Berner1992,Berner2003,Taylor2009,Taylor2011,Ibarra2019,Dahl2020}. The degree to which vascular plants accelerate silicate weathering under a given set of climate conditions is debated, but experiments suggest a biotic weathering enhancement factor of $\approx$2-10 \citep{Moulton1998,Moulton2000,Dahl2020}. 
\begin{center}
\begin{table}[htb!]
\resizebox{\textwidth}{!}{
%\begin{tabular}{||c c c c c||} 
\begin{tabular}{c c c c c} 
 \hline
 Lifespan [Gyr] & Kill mech. & $T_\mathrm{e}$ [K] & $\beta$ & Ref.\\ [0.5ex] 
 \hline
  0.1 & CO$_2$  & n/a & n/a &\citep{Lovelock1982}$^a$ \\ 
 %\hline
  0.9 & CO$_2$ & 13.7 & 0.25$^b$ &\citep{Caldeira1992} \\
 %\hline
 0.5-0.8 & CO$_2$ &  13.7 & 0.25$^b$ & \citep{Franck1999}\\
 %\hline
  0.5 &  CO$_2$&  13.7& 0.25$^b$& \citep{Franck2000}\\
  %\hline
  0.8-1.2& $T$ & 13.7 & 0.25$^b$&  \citep{Lenton2001}$^c$\\
 %\hline
  1.2& CO$_2$ & 13.7 & 0.25$^b$& \citep{Franck2002}\\
 %\hline
 0.5-1.2 & $T$ & 13.7 & 0.25$^b$ & \citep{VonBloh2003}\\
 %\hline
 0.8-1.2 & $T$ & 13.7 & 0.25$^{b}$& \citep{Franck2006}\\
 %\hline
 1.3 & CO$_2$ & 13.7 & 0.3& \citep{Rushby2018}\\
 %\hline
 0.8 & CO$_2$ & 10.9 & 0.5 & \citep{SousaMello2020}\\
 %\hline
 1.0-1.3 & CO$_2$ & 8.0 - 22.2 & 0.5 & \citep{Ozaki2021} \\
 %\hline
 0.9 & CO$_2$ & 17.2 & 0.5 & \citep{Mello2023}$^d$ \\
 \hline
 \textbf{0.5-1.86} & \textbf{$T$} & \textbf{12-48} &  \textbf{0.05-0.91} & \textbf{This work}\\%[1ex] 
 \hline
\end{tabular}
}
\caption{\label{tab:lifespan_studies}\textbf{Estimates of the future lifespan of land plants.} {The first column lists estimates of the future lifespan of vascular land plants from previous work and this study. The second column lists each study's preferred extinction mechanism for land plants: overheating (``$T$'') or CO$_2$ starvation (``CO$_2$''). The third column lists e-folding temperatures for silicate weathering used in each study. The fourth column lists power-law CO$_2$ dependences of silicate weathering used in each study. The final column provides provides references. The bottom row (in bold) summarizes the results and weathering model parameters used in this study.} \\$^a$ \citet{Lovelock1982} assumed land plant extinction below $p$CO$_2$ = 15 Pa (150 ppmv) and did not utilize an explicit weathering model. \\$^b$ These $\beta$ values come from converting a H$^+$ activity power-law \citep{Caldeira1992}, which has a 0.5 power, to the equivalent direct CO$_2$ power law, which reduces the power by a factor of two \citep{Berner1992}. \\$^c$ \citet{Lenton2001} included an additional ``biotic enhancement factor'' in their weathering model that results in somewhat different behavior from the weathering models in the other studies listed. \\$^d$\citet{Mello2023} included a seafloor weathering component with an e-folding temperature of 9.9 K and a CO$_2$ power of 0.3.}
\end{table}
\end{center}

In this study, we apply global-mean models of plant productivity, the carbon cycle, and climate to constrain the lifespan and eventual extinction mechanism of land plants and the species that rely on them (the complex biosphere). For the first time, we consider the weaker temperature dependence and potential for a stronger CO$_2$ dependence of silicate weathering suggested by recent data. We also separately model C$_3$ and C$_4$ plant productivity and their effect on silicate weathering. With these new constraints on silicate weathering, and barring technological intervention or extreme evolutionary adaptation, we find that the complex terrestrial biosphere will be exterminated thermally at temperature and insolation levels approaching moist or runaway greenhouse conditions \citep{Leconte2013,Wolf2014,Wolf2015}. This result emphasizes the importance of carbon cycle parameterization for predicting Earth's far future and underscores the need for further validation with more sophisticated climate models. If life is common beyond Earth, our conclusions may be testable with future observations of biosignatures on extrasolar planets. In the discussion we also touch on potential implications of a longer biosphere lifespan for the prevalence of life in the universe.

\section{Methods}
\subsection{Plant CO$_2$ and temperature limits}
Most previous attempts to estimate the future lifespan of this component of the biosphere have assumed a minimum CO$_2$ for land plants (``CO$_2$ compensation point'') of 10 parts per million by volume (ppmv) of CO$_2$ and a maximum growth temperature of 323 K \citep{Caldeira1992,Franck1999,Franck2000,Lenton2001,Franck2002,VonBloh2003,Franck2006,Rushby2018,SousaMello2020,Ozaki2021,Mello2023}. We argue that both of these choices are too conservative.

Land plant carbon fixation is driven by two main metabolisms known as C$_3$ and C$_4$, responsible for $\approx$70-82\% and $\approx$18-30\% of plant productivity respectively \citep{Francois1998,Raven2008,Blunier2012}. C$_3$ plants, lacking the mechanism C$_4$ plants use to concentrate CO$_2$ near the Rubisco proteins in their cells, have a relatively high CO$_2$ compensation point between approximately 30 and 100 ppmv CO$_2$, depending on plant species and background temperature (under modern O$_2$ conditions) \citep{Collatz1992,Nobel2020}. C$_4$ plants, due to their carbon concentrating mechanism \citep{Raven2008,Edwards2012}, can survive down to much lower CO$_2$ levels of between $\approx$0 and 10 ppmv \citep{Chen1970,Collatz1992,Nobel2020}. As described in section \ref{subsec:productivity}, our default parameter choices for the C$_4$ plant productivity model utilized in this study produce a CO$_2$ compensation point of 2.9 ppmv, which serves as the minimum CO$_2$ level required for plant growth in this study. This may be a conservative choice, since several plants are known to have CO$_2$ compensation points below this limit \citep{Chen1970}.

\begin{figure*}[htb!]
    \centering
    \makebox[\textwidth][c]{\includegraphics[width=270pt,keepaspectratio]{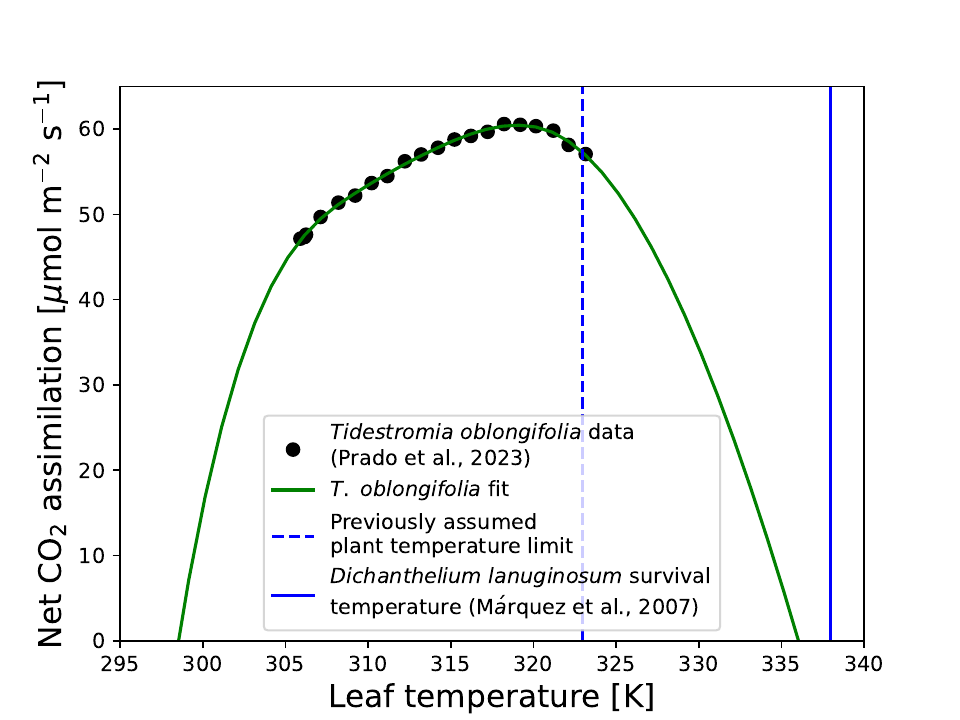}}
    \caption{\textbf{Comparing \textit{Tidestromia oblongiformia} and \textit{Dichanthelium lanuginosum} temperature data to previously assumed plant growth temperature limit.} Black circles show observations of the rate of photosynthesis as a function of temperature for \textit{T. oblongiformia} \citep{Prado2023}, which has the highest optimal growth temperature observed in any vascular plant \citep{Bjoerkman1972}. The green curve is a cubic spline fit to the \textit{T. oblongiformia} data, providing an estimate of 336 K for the extremophile's upper temperature limit. The dashed blue line shows the 323 K upper temperature limit for land plants that previous studies have used, which is contradicted by the observed ability of \textit{T. oblongiformia} to photosynthesize at a high rate at that temperature. The solid blue line is at 338 K, the maximum temperature at which the panic grass \textit{Dichanthelium lanuginosum} has been shown to be able to survive repeated longterm exposure when inoculated with a symbiotic virus-infected root fungus \citep{Redman2002,Marquez2007}. That is the temperature limit for land plants that we assume in this study.}
    
    \label{fig:death_valley}
\end{figure*}

As noted before, the upper temperature limit for land plants in studies of the future lifespan of the complex terrestrial biosphere is frequently taken to be 323~K (50~C) \citep[e.g.][]{Caldeira1992}. However, a number of plants living in desert and hydrothermal environments can survive and photosynthesize through periodic exposure to temperatures at and above this putative limit, and plant temperature optima tend to mirror the maximum temperatures encountered in their local environments \citep{Prado2023}. These factors suggest the range of observed thermal tolerances for plants may simply reflect the range of environments present on Earth today. As an example of a plant that pushes up against the 323 K boundary, field observations of the C$_4$ shrub \textit{Hammada scoparia} in the Negev desert suggest a seasonal photosynthetic optimum temperature of 313~K (40~C), with photosynthesis remaining above zero up to a temperature of 333~K (60~C) \citep{Lange1974}. A few plants display photosynthetic temperature optima even higher than \textit{H. scoparia} \citep[][Data S3]{Prado2023}. One of the most extreme known examples is the thermophilic C$_4$ desert perennial \textit{Tidestromia oblongifolia}, which reaches a maximum photosynthesis rate at 320~K (47~C) \citep{Bjoerkman1972} and grows exponentially during Death Valley summers that regularly subject plants to temperatures of between 46 and 50~C \citep{Prado2023}. Although \textit{T. oblongifolia}'s upper temperature limit for photosynthesis does not appear to have been recorded, its optimum temperature is 7~K higher than that of \textit{H. scoparia}, suggesting it may be able to photosynthesize several degrees above \textit{H. scoparia}'s 333~K limit. Extrapolation with a simple cubic spline fitted to observations from \citet{Prado2023} of \textit{T. oblongifolia}'s photosynthesis rate between 306 K (33~C) and 323 K (50~C) suggests the species could potentially photosynthesize (likely with significant damage) up to 336~K (63~C) (see Fig. \ref{fig:death_valley}). Regardless of the exact value of its upper temperature limit, the fact that \textit{T. oblongiformia} operates with near-peak photosynthetic rates at 323~K seems to invalidate that temperature as a fundamental threshold for plant growth. 

The upper temperature limit for land plants can be estimated from the highest temperature they have been observed to survive \citep{Clarke2014}. Although a thermal limit for eukaryotic organisms in general of 333~K is often cited \citep{Tansey1972}, the temperature record for land plants appears to be held by \textit{Dichanthelium lanuginosum}, a C$_3$ grass growing in geothermal settings that can remain healthy through weeks of daily 10-hour-long exposures to rhizosphere temperatures of 338~K (when colonized by the fungus \textit{Curvularia protuberata} while the fungus is infected with Curvularia thermal tolerance virus) \citep{Brown1975,Redman2002,Marquez2007,Clarke2014}. A similar degree of thermotolerance was conferred on tomato plants \citep{Marquez2007}, watermelons, and wheat \citep{Pennisi2003} that were inoculated with the virus-infected fungus, suggesting the mechanism mediating the effect in the grass may be generally applicable. At face value, the possibility of plant survival at 338~K seems implausible, in particular since enzymes necessary to maintain Rubisco activity have been found to rapidly denature and become inactive below 323~K \citep{Salvucci2001}, which is thought to be one of the primary drivers of photosynthetic inhibition under temperature stress \citep{Salvucci2004,Scafaro2023}. Nonetheless, there is significant variation in thermotolerance of these enzymes among clades \citep{Salvucci2004,Shivhare2017,Perkins2021}, with more thermotolerant forms directly resulting in higher optimal growth temperatures for plants \citep{Salvucci2004,Scafaro2019}. Furthermore, cyanobacteria in hot springs utilizing the same basic photosynthetic machinery have been found to operate at temperatures as high as 347~K (74 C) \citep{Miller2013}, demonstrating there is no fundamental ceiling for photosynthesis as a mechanism below that temperature. Given the apparent ability of plants in certain conditions to survive repeated exposures of 338~K, in this paper we take this as the upper temperature limit for plants, while including some calculations and remarks on the implications for our results if the tradition plant growth limit of 323~K is retained.

\subsection{Weathering model}\label{subsec:weathering}
We adopt a standard expression \citep{Walker1981, Abbot2016} for silicate weathering as a function of surface temperature ($T$) and soil $p$CO$_2$ ($p_\mathrm{soil}$):
\begin{align}\label{eqn:weathering}
\frac{W}{W_\mathrm{ref}} &= \frac{V}{V_\mathrm{ref}} = \mathrm{exp}({\frac{T-T_\mathrm{ref}}{T_\mathrm{e}}})(\frac{p_\mathrm{soil}}{p_\mathrm{soil,ref}})^{\beta}
\end{align}
where $W$ [mol yr$^{-1}$] is the global silicate weathering rate; $T_\mathrm{e}$ [K] is the e-folding temperature for weathering; $\beta$ is the exponent for the power-law dependence on soil CO$_2$; $T$ [K] and $p_\mathrm{soil}$ [bar] are defined as above; $V_\mathrm{ref}=W_\mathrm{ref}=7.5\times10^{12}$ moles of CO$_2$ per year is modern Earth's approximate CO$_2$ outgassing rate \citep{HaqqMisra2016}; $T_\mathrm{ref}=288$ K is the pre-industrial Earth's global-mean temperature; $p_\mathrm{soil,ref}=2.8\times10^{-3}$ bars (since the modern biota maintains soil CO$_2$ at approximately an order of magnitude higher level than atmospheric CO$_2$ \citep{Volk1987}). We test the full range of $T_\mathrm{e}$ and $\beta$ values described in Section \ref{sec:intro}, using $T_\mathrm{e}\in$[12.1, 48] K and $\beta\in$[0.05, 0.91] for our calculations. We note that, if global runoff is linearly dependent on temperature as is often assumed \citep{D.S.Abbot2012,Abbot2016,Graham2020,Coy2022}, then Taylor expanding the above equation with respect to $T$ and keeping the linear term produces a model equivalent to a global-mean version of the hydrologically-regulated weathering model of \citet{Maher2014} (MAC) in the runoff-limited regime \citep{Graham2020}, such that simulations with small $T_\mathrm{e}$ are equivalent to the MAC model. If the coupling of climate to weathering emerges from the relationship between runoff and global surface temperature rather than silicate dissolution kinetics \citep{Godsey2009,Maher2014,Graham2020}, then we estimate $T_\mathrm{e}\approx$15--44~K from modeled temperature sensitivities of high and low latitude rivers \citep{Manabe2004,Maher2014}. So, our results should still apply even if the MAC model is a more mechanistically accurate model of weathering than the exponential formulation we are using here. 

Taking a logarithm of equation \ref{eqn:weathering} shows that for simulations with the reference outgassing rate ($V=V_\mathrm{ref}$), the trajectory is determined by the product of $\beta$ and $T_\mathrm{e}$, i.e. simulations with $T_\mathrm{e}=20$ K and $\beta=0.25$ produce the same trajectories as simulations with $T_\mathrm{e}=10$ K and $\beta=0.5$:
\[
\log\big(\frac{V}{V_\mathrm{ref}}\big) = \frac{T - T_\mathrm{ref}}{T_\mathrm{e}} + \beta\log\big(\frac{p_\mathrm{soil}(T,p_\mathrm{atm})}{p_\mathrm{soil,ref}}\big)\]
\[
\frac{T_\mathrm{ref} - T}{\log\big(\frac{p_\mathrm{soil}(T,p_\mathrm{atm})}{p_\mathrm{soil,ref}}\big)} =  T_\mathrm{e}\beta
\]
so we label results for widely varying $T_\mathrm{e}$ and $\beta$ by the product of the two weathering parameters ($\beta\times T_\mathrm{e}$) in some of the results below.

\subsection{Soil CO$_2$ model}
Soil $p$CO$_2$ ($p_\mathrm{soil}$) is maintained above its atmospheric level by below-ground respiration of plant and fungal organic matter; we parameterize this effect as a function of atmospheric $p$CO$_2$ ($p_\mathrm{atm}$) and NPP following \citet{Volk1987}: 
\begin{align}\label{eqn:soil_pco2}
    \frac{p_\mathrm{soil}}{p_\mathrm{soil,ref}} &= R_\mathrm{NPP}(1-\frac{p_\mathrm{atm,ref}}{p_\mathrm{soil,ref}}) + \frac{p_\mathrm{atm}}{p_\mathrm{soil,ref}}
\end{align}
where $p_\mathrm{atm}$ [bars] is the atmospheric $p$CO$_2$; $p_\mathrm{atm,ref}=2.8\times10^{-4}$ bars CO$_2$ is the pre-industrial atmospheric CO$_2$ partial pressure; $R_\mathrm{NPP}$ is the total net primary productivity relative to modern; and other variables are defined as above. We note that if rates of organic decay are strongly temperature-dependent, then $p_\mathrm{soil}$ might be sustained at values significantly higher or lower than those given in our study. 

\subsection{C$_3$ and C$_4$ plant productivity models}\label{subsec:productivity}
Relative net primary productivity ($R_\mathrm{NPP}$) is taken to be a linear sum of the NPP values predicted by idealized C$_3$ and C$_4$ photosynthesis models, relative to model output for modern pre-industrial $T$ and $p_\mathrm{atm}$ and normalized to estimates of the modern fractions of organic carbon fixation by each pathway:
\begin{eqnarray*}
    %\begin{align}
        R_\mathrm{NPP}(T,p_\mathrm{atm}) &= f_\mathrm{C3,modern} R_\mathrm{NPP,C3}(T,p_\mathrm{atm}) + f_\mathrm{C4,modern} R_\mathrm{NPP,C4}(T,p_\mathrm{atm})\\
        &= 0.725 \frac{NPP_\mathrm{C3}(T,p_\mathrm{atm})}{NPP_\mathrm{C3}(288 \mathrm{ K},280\times10^{-6} \mathrm{ bar})} + 0.275\frac{NPP_\mathrm{C4}(T,p_\mathrm{atm})}{NPP_\mathrm{C4}(288 \mathrm{ K},280\times10^{-6} \mathrm{ bar})}        
    %\end{align}
\end{eqnarray*}
where $f_\mathrm{C3,modern}=0.725$ and $f_\mathrm{C4,modern}=0.275$ are the modern fractions of organic carbon fixation by C$_3$ and C$_4$ plants respectively according to models of glacial-interglacial changes in global primary productivity during the current ice age \citep{Francois1998,Blunier2012} and $NPP_\mathrm{C3}(T,p_\mathrm{atm})$ and $NPP_\mathrm{C4}(T,p_\mathrm{atm})$ are C$_3$ and C$_4$ productivities in $\mu$mol s$^{-1}$ per unit leaf area from biochemical models of photosynthetic CO$_2$ assimilation for each plant type. 

For C$_3$ plants, we apply a version of the biochemical leaf photosynthesis model from \citet{Farquhar1980}. We use temperature response functions from \citet{Bernacchi2001}. For simplicity, we assume C$_3$ photosynthesis is always limited by Rubisco carboxylation (as opposed to other potentially limiting processes like RuBP regeneration or triose-phosphate utilization \citep{Farquhar1980}), since photosynthesis is commonly limited by Rubisco's kinetics in nature \citep{Rogers2000,Bernacchi2001}:
\begin{equation}
        \label{eqn:c3_productivity}NPP_\mathrm{C3}(T,p_\mathrm{atm}) = (1-\frac{\Gamma_*(T)}{C_\mathrm{i}(T,p_\mathrm{atm})}) \frac{ V_\mathrm{c,max}(T) C_\mathrm{i}(T,p_\mathrm{atm})}{(C_\mathrm{i}(T,p_\mathrm{atm}) + K_\mathrm{c}(T)(1 + \frac{O}{K_\mathrm{o}(T)}))} - R_\mathrm{d}(T)
\end{equation}

where $\Gamma_*(T)=\exp(19.02 - 37.83/(RT))$ is the CO$_2$ compensation point for C$_3$ plants in the absence of respiration, given as a CO$_2$ concentration in air [$\mu$mol mol$^{-1}$]; $C_i(T,p_\mathrm{atm})$ is intercellular CO$_2$ concentration [$\mu$mol mol] maintained by exchange of air and water through leaf stomata; $V_\mathrm{c,max}(T)=V_\mathrm{c,max}(298\mathrm{ K})\exp(26.35 - 65.33/(RT))$ is the maximum rate of carboxylation by Rubisco [$\mu$mol m$^{-1}$ s$^{-1}$] at a given leaf temperature $T$ (assumed equal to global-mean surface temperature in our models); $V_\mathrm{c,max}(298)=57.05$ $\mu$mol m$^{-1}$ s$^{-1}$ is maximum carboxylation rate at 298 K for \textit{Pinus talleda} \citep{Lin2012} (a randomly chosen C$_3$ plant);  $K_\mathrm{c}=\exp(38.05 - 79.43/(RT))$ [$\mu$mol mol$^{-1}$] and $K_\mathrm{O}=\exp(20.30 - 36.38/(RT))$ [mmol mol$^{-1}$] are Rubisco's temperature-dependent Michaelis–Menten constants for CO$_2$ and O$_2$ respectively; $O=0.21\times10^{3}$ is the atmospheric concentration of oxygen [mmol mol$^{-1}$]; and $R_\mathrm{d}=\exp(18.72 - 46.39/(R*T))$ is mitochondrial respiration in illuminated conditions [$\mu$mol m$^{-2}$ s$^{-1}$]. $C_\mathrm{i}$ is calculated with a simple stomatal conductance model drawn from the corrected version of \citet{Medlyn2011,Medlyn2012}. We assume a constant humidity deficit between air and intercellular spaces of leaves and a constant marginal water cost of carbon ($\lambda$) and we apply parameter values for the Duke pine (a C$_3$ plant) from Lin et al. \citep{Lin2012}:
\begin{equation}
    C_\mathrm{i} = C_\mathrm{a} \frac{g_1}{g_1+\sqrt{D}}
\end{equation}
where $C_\mathrm{a}$ is the CO$_2$ concentration in air [$\mu$mol mol$^{-1}$]; $D=2.5$ kPa is an arbitrary assumed constant humidity deficit between the saturated intercellular space of the leaves and the surrounding air (results are not sensitive to its precise value); and $$g_1=10.96\sqrt{\frac{\Gamma_*(T)}{\Gamma_*(28.1)}}$$ is a fitting parameter that is proportional to $\sqrt(\Gamma_*(T)\lambda)$ (see equation (11) in \citet{Medlyn2011}), scaled with a constant $\lambda$ to the value of $g_1=10.96$ at 28.1 K given for the Duke pine in \citet{Lin2012}.

We use a different productivity model for C$_4$ plants which accounts for the fact that C$_4$ photosynthesis is nearly insensitive to CO$_2$ at modern-Earth-like levels due to efficient spatial concentration of CO$_2$ around Rubisco in bundle sheaths, but becomes linearly dependent on CO$_2$ at low concentrations when the C$_4$ plant's bundle sheaths are subsaturated with CO$_2$. The transition between these regimes usually occurs around 100 ppmv CO$_2$ \citep{Collatz1992,Chen1994}. We handle this by calculating both CO$_2$-saturated and CO$_2$-subsaturated rates and assuming that the slower of the two at any given $T$ and $p_\mathrm{atm}$ operates as the limiting factor on photosynthesis (we always assume fully light-saturated photosynthesis because we cannot account for spatial variations in light and because light limitation should become less and less of an issue for plants under a gradually brightening sun): 
%\begin{linenomath}
\begin{eqnarray}\label{eqn:c4_photosynthesis} 
        NPP_\mathrm{C4}(T,p_\mathrm{atm}) = -R_\mathrm{d}(T) + 
        \min\begin{cases}
        A_\mathrm{CO_2} &= 1.8\times10^4V(T)\frac{p_\mathrm{i}(p_\mathrm{atm})}{p_\mathrm{tot}}\\
        A_\mathrm{Rubisco} &= V(T)
        \end{cases}
\end{eqnarray}
%\end{linenomath}
where $V(T)=V_\mathrm{max}2^{H(T)}$ is the temperature-dependent rate of light- and CO$_2$-saturated Rubisco carboxylation [$\mu$mol m$^{-2}$ s$^{-1}$]; $V_\mathrm{max}=39$ $\mu$mol m$^{-2}$ s$^{-1}$; 
$H(T)={\frac{(T-298.15)/10}{(1+\exp(0.3(286.15-T)))(1+\exp(0.3(T-309.15)))}}$;
$p_\mathrm{i}=0.4p_\mathrm{atm}$ [bars] is the intercellular CO$_2$ partial pressure, taken to be 40$\%$ of the ambient atmospheric CO$_2$ as a first approximation based on measurements suggesting an approximately constant offset between ambient and intercellular CO$_2$  \citep{Wong1979,Jacobs1994}; $p_\mathrm{tot}=1$ bar is the total atmospheric pressure; and as suggested in \citet{Collatz1992}, Rubisco-limited photosynthesis ($A_\mathrm{Rubisco}$), CO$_2$-subsaturated photosynthesis ($A_\mathrm{CO_2}$), and plant respiration ($R_\mathrm{d}$=0.021$V(T)$) are all assumed to be linearly related \citep{Cox1998}. This simple formulation allows us to explicitly calculate the CO$_2$ where C$_4$ plants transition to CO$_2$ subsaturation in our model,  $p_\mathrm{atm}=1/(1.8\times10^4)/0.4=1.4\times10^{-4}$ bars, or 140 ppmv CO$_2$, as well as the CO$_2$ compensation point for C$_4$ plants, $p_\mathrm{atm}=0.021/(1.8\times10^4)/0.4=2.9\times10^{-6}$ bars, 2.9 ppmv CO$_2$. The transition between CO$_2$ saturation and subsaturation for C$_4$ plants is visible in Fig. \ref{fig:carbon_cycle_contours} as a discontinuity in the slope of all carbon cycle trajectories at 140 ppmv CO$_2$.

To examine whether the productivity model produces reasonable responses to changes in climate, we compared our results against relative productivity values inferred from ice cores for glacial-interglacial changes in CO$_2$ and temperature over the past 800,000 years \citep{Yang2022}. Ocean productivity is estimated to have been within 20$\%$ of its modern value over the last 400,000 years \citep{Blunier2012}, suggesting the large productivity changes inferred by \citet{Yang2022} are largely attributable to the land biosphere, facilitating comparison with our model. \citet{Yang2022} found glacial productivity minima between 55 and 87\% of modern for conditions $\approx5$ K cooler than today with CO$_2\approx200$ ppmv. Our model produces an equivalent result: with $T=283$ K and $p_\mathrm{atm}=2\times10^{-4}$ bar, $R_\mathrm{NPP}=0.56$, which falls within the inferred range and suggests our model at least qualitatively captures the global response of plant productivity to changes in climate.

\subsection{Climate model}
The climate model assumes global-mean radiation balance between the planet's absorbed solar radiation and outgoing longwave radiation:
\begin{equation}
    ASR = OLR
\end{equation}
where $ASR$ is absorbed solar radiation and $OLR$ is outgoing longwave radiation. $ASR$ is defined by:
\begin{equation}\label{eqn:climate}
    ASR = (1-a)\frac{S}{4}
\end{equation}
where $a=0.29$ is the planetary Bond albedo \citep{Stephens2015}, assumed constant in our model due to uncertainty in the magnitude and direction of the cloud feedbacks that dominate planetary albedo response to changes in CO$_2$ and insolation \citep{Leconte2013,Wolf2014,Wolf2015,Popp2016,Wolf2018}, and $S$ is the top-of-atmosphere substellar insolation, divided by a factor of 4 to account for the distribution of radiation intercepted by Earth's circular silhouette across its spherical surface. $S$ is related to time in the future by interpolation of luminosity vs. time data from a standard stellar evolution calculation presented in \citet{Bahcall2001}. $OLR$ is defined using the fit to H$_2$O-saturated radiative-convective column models presented in \citet{Caldeira1992} with an additional tuning factor of -18 W m$^{-2}$ to reproduce Earth's climate with the modern albedo $a=0.29$ at the modern insolation of $S=1361$ W m$^{-2}$: 
\begin{equation}
    OLR = \sigma T_\mathrm{eff}^4  - 18  
\end{equation}
where $\sigma$ is 5.67$\times10^{-8}$ W m$^{-2}$ K$^{-4}$ and $T_\mathrm{eff}$ is Earth's effective radiating temperature. $T_\mathrm{eff}$ is defined by:
\begin{equation}
    T_\mathrm{eff} = T - \Delta T
\end{equation}
where $T$ is the global-mean surface temperature and $\Delta T$ is the warming due to the greenhouse effect, a polynomial fit to output from radiative-convective column model simulations provided by \citet{Caldeira1992}:
\begin{equation}
    \Delta T = 815.17 + \frac{4.895\times10^7}{T^{2}} -\frac{3.9787\times10^5}{T} - \frac{6.7084}{(\log(p_\mathrm{atm}))^{2}} + \frac{73.221}{\log(p_\mathrm{atm})} - \frac{3.0882\times10^4}{T\log({p_\mathrm{atm}})}
\end{equation}
where $T$ is Earth's global-mean surface temperature in Kelvin and $p_\mathrm{atm}$ is the atmospheric CO$_2$ partial pressure in bars.

\section{Results}
\begin{figure*}[htb!]
    \centering
    \makebox[\textwidth][c]{\includegraphics[width=200pt,keepaspectratio]{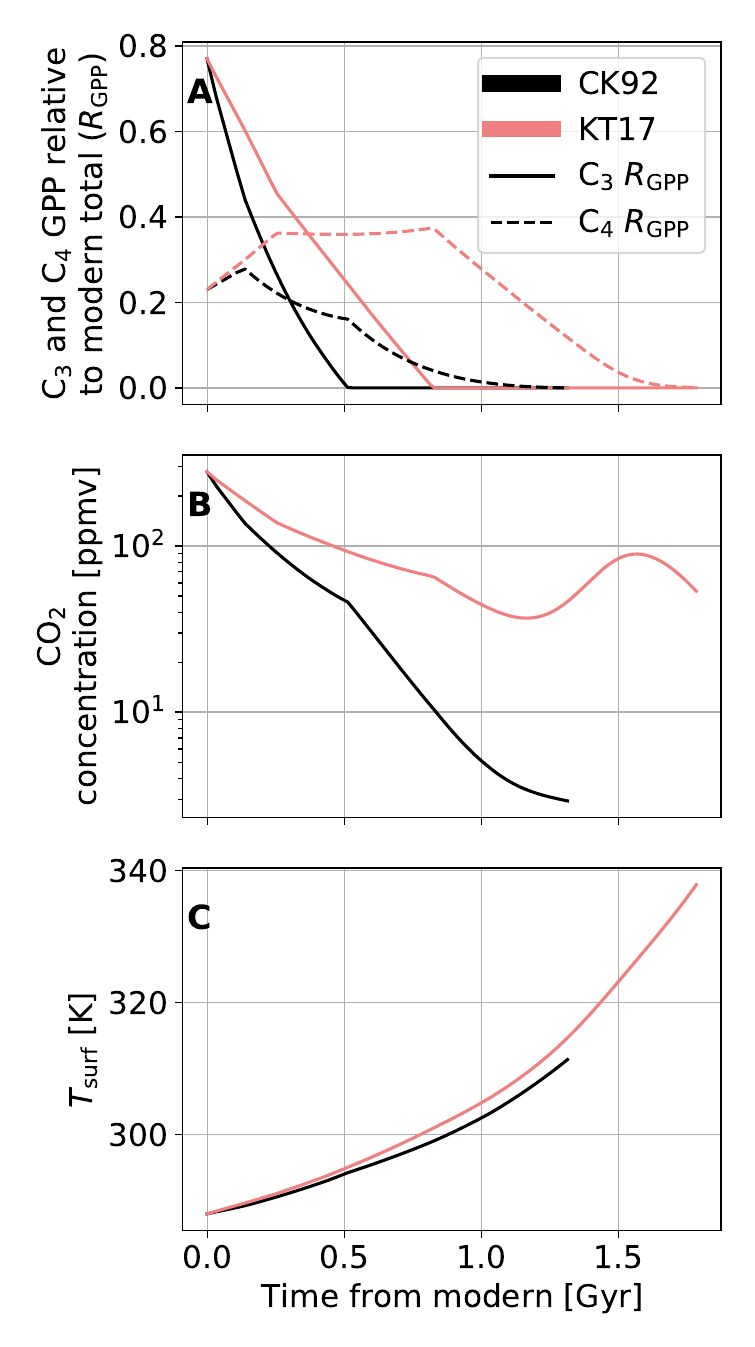}}
    \caption{\textbf{Representative future climate and biosphere trajectories.} (\textbf{A}) Trajectories of C$_3$ (solid) and C$_4$ (dashed) plant productivity relative to total modern productivity. (\textbf{B}) Atmospheric CO$_2$ concentration (ppmv). (\textbf{C}) Surface temperature ($T_\mathrm{surf}$ in Kelvins). Results are shown with weathering parameters from \citet{Caldeira1992} (black, $T_e=13.7$ K, $\beta=0.25$) and from Krissansen-Totton and Catling \citet{KrissansenTotton2017} (pink, $T_e=31.0$ K, $\beta=0.41$).}
    
    \label{fig:standard_fiducial_comparison}
\end{figure*}

First, we compare future biosphere trajectories with parameterizations that represent the two classes of land plant extinction---CO$_2$ starvation and overheating. The parameterization leading to extinction from inadequate carbon dioxide follows \citet{Caldeira1992} (henceforth CK92), who used $T_\mathrm{e}=13.7$ K and $\beta=0.25$. The parameterization leading to overheating relies on the study of \citet{KrissansenTotton2017} (henceforth KT17), who constrained the distributions of $T_\mathrm{e}$ and $\beta$ values that are consistent with paleoclimate proxy records. We use their median $T_\mathrm{e}=31$ K and $\beta=0.41$ values, which were drawn from a Michaelis-Menten weathering formulation often used to approximate the impact of vascular plants on weathering rates \citep{Volk1987}. Together, these two simulations display the main qualitative properties of the two classes of land plant extinction trajectories.

The KT17 parameters result in a longer future lifespan of terrestrial plants relative to the CK92 parameters (1.8 and 1.3 Gyr, respectively, Fig.\,\ref{fig:standard_fiducial_comparison}A). The stronger dependence of the silicate weathering rate on temperature (smaller $T_\mathrm{e}$) with CK92 parameters leads to a stronger response to increased luminosity, monotonically decreasing atmospheric CO$_2$ concentration and ultimately leading to plant extinction by CO$_2$ starvation. In contrast, the weaker temperature dependence {(larger $T_\mathrm{e}$)} with the KT17 parameters results in higher atmospheric CO$_2$ levels for a given solar luminosity (Fig.\,\ref{fig:standard_fiducial_comparison}B), avoidance of {plant} starvation, and ultimate extinction by overheating. The 1.3 Gyr lifespan that we calculate with the CK92 parameterization is substantially longer than the 0.9 Gyr calculated in the original study because in our model C$_4$ plants have a {lower} compensation threshold (2.9 ppmv, rather than 10 ppmv). With a threshold of 10 ppmv, as assumed in previous biosphere lifespan studies, the difference between the CK92 and KT17 lifespans is even larger: $\approx$0.9 Gyr as opposed to $\approx$1.8 Gyr.

In both simulations, C$_3$ plant productivity monotonically decreases, but in the CK92 simulation {C$_3$ extinction occurs} after 0.5 Gyr, while in the KT17 simulation they last for 0.8 Gyr (Fig. \ref{fig:standard_fiducial_comparison}A), as the CO$_2$ decrease with increasing luminosity is less pronounced. In both cases, the fall in C$_3$ productivity reduces soil respiration, {decreasing} soil CO$_2$ ($p_\mathrm{soil}$ in Eq.~\ref{eqn:weathering} and \ref{eqn:soil_pco2}). This reduces the biotic acceleration of weathering and slows the drawdown of CO$_2$ with increasing solar luminosity, extending the lifespan of C$_3$ plants by delaying starvation. This is reflected in the shallow slope of the CO$_2$ vs. time curve for each simulation when C$_3$ plants exist (Fig.\,\ref{fig:standard_fiducial_comparison}B). The delayed CO$_2$ decrease also sustains C$_4$ productivity at high levels for long periods in both simulations (Fig.\,\ref{fig:standard_fiducial_comparison}A). 

Initially, {during the period} when C$_4$ {plants operate without} CO$_2$-{limitation, their} productivity is insensitive to {changes in} atmospheric {composition}. Over this interval, increasing temperature leads to climbing C$_4$ productivity. {Only after 0.12 (CK92 parameters) or 0.25 Gyr (KT17 parameters) does} CO$_2$ limitation of photosynthesis {begin to outweigh} temperature fertilization {for C$_4$ plants in our simulations}. With the CK92 parameters, C$_4$ productivity begins to fall {after 0.12 Gyr}, whereas with the KT17 parameters, C$_4$ productivity stabilizes {at 0.25 Gyr} at a value {for} which the reduction from gradually decreasing CO$_2$ is approximately balanced by the increase from warming temperatures. C$_4$ productivity remains at this high value until C$_3$ plants die off, at which point {atmospheric composition begins to change more rapidly} because the C$_3$ productivity decrease can no longer reduce $p_\mathrm{soil}$ and {alter} the weathering rate. This causes C$_4$ productivity to finally start declining 0.8 Gyr from now, falling below its present value only after 1.1 Gyr.

As C$_4$ productivity falls after 0.8 Gyr in the KT17 simulation, the CO$_2$ decrease is initially rapid but slows with time, until around 1.1 Ga it halts and then reverses. This is due to the same fundamental mechanism as the initial decelerated loss of CO$_2$ from decreasing C$_3$ productivity during the first few hundred million years described above. As C$_4$ productivity falls, root respiration slows ($p_\mathrm{soil}$ approaches $p_\mathrm{atm}$ in Eq.\ref{eqn:soil_pco2}), necessitating higher temperature and atmospheric CO$_2$ {levels} to {maintain} a given weathering rate. With the large $T_\mathrm{e}$ in KT17, an atmospheric CO$_2$ \textit{increase} becomes necessary to offset the decline in weathering rate from {declining} productivity, and only after temperatures climb past 327 K at 1.6 Gyr does the temperature effect on weathering begin to dominate and resume net CO$_2$ drawdown. In contrast to the rolling trajectory {displayed by} the KT17 simulation, CO$_2$ in the CK92 simulation falls faster and monotonically with time, {leading to cooler conditions}. The {smaller values of $T_\mathrm{e}$ and $\beta$ in the CK92 parameterization} require a smaller temperature change and a larger CO$_2$ change to sustain a given weathering rate upon a change in solar luminosity.

\begin{figure*}[htb!]
    \centering
    \makebox[\textwidth][c]{\includegraphics[width=500pt,keepaspectratio]{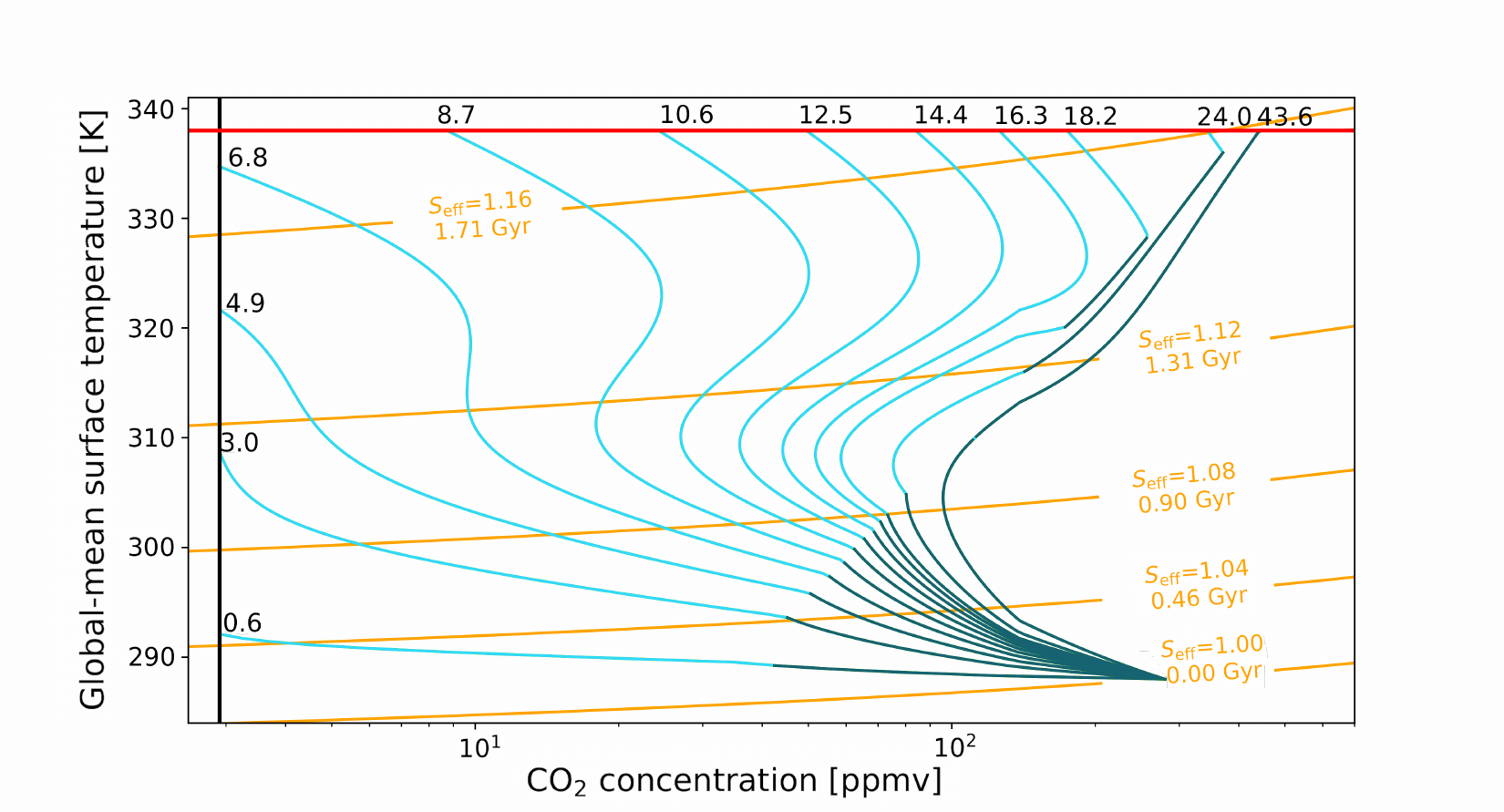}}
    \caption{\textbf{Climate trajectories in CO$_2$ (ppmv)-temperature (Kelvins) space with modern outgassing for different values of $T_\mathrm{e}\times\beta$ (Kelvins).} Blue curves (both light and dark) show temperature-CO$_2$ combinations that produce carbon cycle equilibrium at modern outgassing for different combinations of weathering parameters. Each blue curve is labeled by its corresponding $T_\mathrm{e}\times\beta$ value. Dark blue portions of the curves have both C$_3$ and C$_4$ plants, while light blue portions have only C$_4$ plants. The horizontal red line marks our assumed temperature threshold for land plants, 338 K \citep{Redman2002,Marquez2007,Clarke2014}. The vertical black line marks the CO$_2$ compensation point for C$_4$ plants that emerges from our photosynthesis model, 2.9 ppmv. Orange contours show temperature-CO$_2$ combinations that produce top-of-atmosphere radiative equilibrium at a given effective insolation ($S_\mathrm{eff}$) and corresponding time in the future, so time and insolation increase from the bottom-right to the top-left of the figure.}
    \label{fig:carbon_cycle_contours}
\end{figure*}

For modern outgassing, model trajectories are determined by the product of $\beta$ and $T_\mathrm{e}$ (K), rather than each parameter independently (Fig.\,\ref{fig:carbon_cycle_contours}; section \ref{subsec:weathering}). For $\beta\times T_\mathrm{e}\in[0.6,43.6]$ within estimates of the plausible range \citep{KrissansenTotton2017,Deng2022}, we find substantial variation not only in timing of biosphere demise, but also in kill mechanism. For relatively high temperature sensitivity (low $T_\mathrm{e}$) and/or low CO$_2$ sensitivity (low $\beta$), like CK92, the biosphere dies by CO$_2$ starvation after monotonic or nearly monotonic drawdown of CO$_2$ to 2.9 ppmv (the C$_4$ compensation point; Eq. \ref{eqn:c4_photosynthesis}). For settings like KT17, with high $T_\mathrm{e}$ and/or high $\beta$, the biosphere overheats at 338 K after a non-monotonic atmospheric evolution driven by the shifting balance between the contrasting effects of increasing temperature and decreasing soil CO$_2$ on the weathering rate, as described above.

For most parameter values, C$_3$ plants die off permanently from CO$_2$ starvation, leaving behind biospheres entirely consisting of C$_4$ plants. The CO$_2$ at which this occurs varies among the trajectories because the C$_3$ CO$_2$ compensation point is temperature-dependent (Eq.\ref{eqn:c3_productivity}), with lower temperatures allowing C$_3$ plants to persist to lower CO$_2$ levels \citep{Farquhar1980}. In the simulations with $T_\mathrm{e}\times\beta$=18.2 K and 24.0 K, CO$_2$ eventually reaches high enough levels during the increasing phase that C$_3$ plants are able to re-emerge far in the future, delaying the onset of the CO$_2$ decrease at the end of each trajectory. In the simulation with $T_\mathrm{e}\times\beta=43.6$ K, CO$_2$ only briefly falls to levels low enough for C$_3$ plants to disappear, so they remain present almost continuously until the biosphere overheats at a CO$_2$ of 440 ppmv with no final CO$_2$ decrease. In this limit, the silicate weathering feedback is almost entirely mediated by $p_\mathrm{soil}$, which is in turn determined largely by Net Primary Productivity (NPP), so that a balanced carbon cycle requires maintaining near-modern NPP levels throughout the simulation, which is why CO$_2$ remains high. 
\begin{figure*}[htb!]
    \centering
    \makebox[\textwidth][c]{\includegraphics[width=500pt,keepaspectratio]{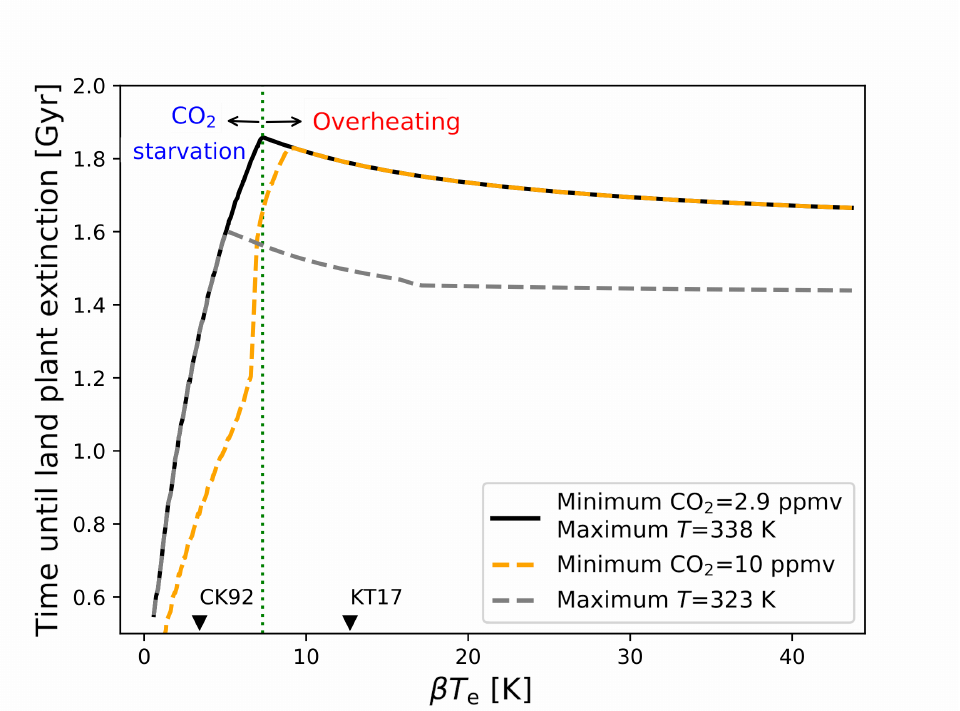}}
    \caption{\textbf{Future lifespan of the complex biosphere [Gyr] as a function of $T_\mathrm{e}\times\beta$ [K] with modern CO$_2$ outgassing.} The black curve records lifespans with our 338 K temperature threshold and the 2.9 ppmv CO$_2$ threshold that emerges from our C$_4$ photosynthesis model. The vertical green dotted line separates the trajectories that end in CO$_2$ starvation for land plants from the trajectories that end in overheating. The grey dashed line shows the time required for our model to reach either our standard CO$_2$ threshold or 323 K, the temperature threshold used by prior studies. The orange dashed line shows the time required for our model to reach either our standard temperature threshold or 10 ppmv CO$_2$, the CO$_2$ minimum assumed by most prior studies.}
    \label{fig:Te_beta_lifespan2}
\end{figure*}

In the CO$_2$ starvation regime, small variations in $T_\mathrm{e}$ and $\beta$ lead to large changes in expected lifespan (Fig.\,\ref{fig:Te_beta_lifespan2}), with increases in $\beta$ or $T_\mathrm{e}$ both increasing lifespan by raising the equilibrium CO$_2$ at a given insolation, delaying starvation. With our climate model, lifespans in the CO$_2$ starvation regime range from 0.55 Gyr at $T_\mathrm{e}\times\beta=0.6$ to a maximum lifespan of 1.86 Gyr at $T_\mathrm{e}\times\beta=7.3$, which marks the transition from CO$_2$ starvation to overheating. At $T_\mathrm{e}\times\beta>7.3$, the plant lifespan shortens with increasing $T_\mathrm{e}\times\beta$ because elevated CO$_2$ causes the planet to reach the 338 K threshold earlier, falling from 1.86 Gyr at $T_\mathrm{e}\times\beta=7.3$ to 1.66 Gyr at $T_\mathrm{e}\times\beta=43.6$ K. A CO$_2$ compensation point of 10 ppmv (orange dashed curve, Fig.\,\ref{fig:Te_beta_lifespan2}) or upper temperature threshold of 323 K (grey dashed curve, Fig.\,\ref{fig:Te_beta_lifespan2}) can each shorten lifespans by a few hundred million years.

So far, all presented simulations have assumed outgassing and land fraction equal to modern values, but in the far future geodynamic effects may reduce Earth's CO$_2$ outgassing rate and drive continental growth  \citep{Franck2000}. Both changes would reduce the equilibrium CO$_2$ of the atmosphere, hastening CO$_2$ starvation \citep{Franck1999,Franck2000,Lenton2001}. To test this possibility, we calculate the lifespan of land plants under different constant outgassing rates. In our zero-dimensional carbon cycle model, a given reduction in outgassing is equivalent to an increase in continental area or weatherability by the same factor at constant outgassing.

\begin{figure*}[htb!]
    \centering
    \makebox[\textwidth][c]{\includegraphics[width=350pt,keepaspectratio]{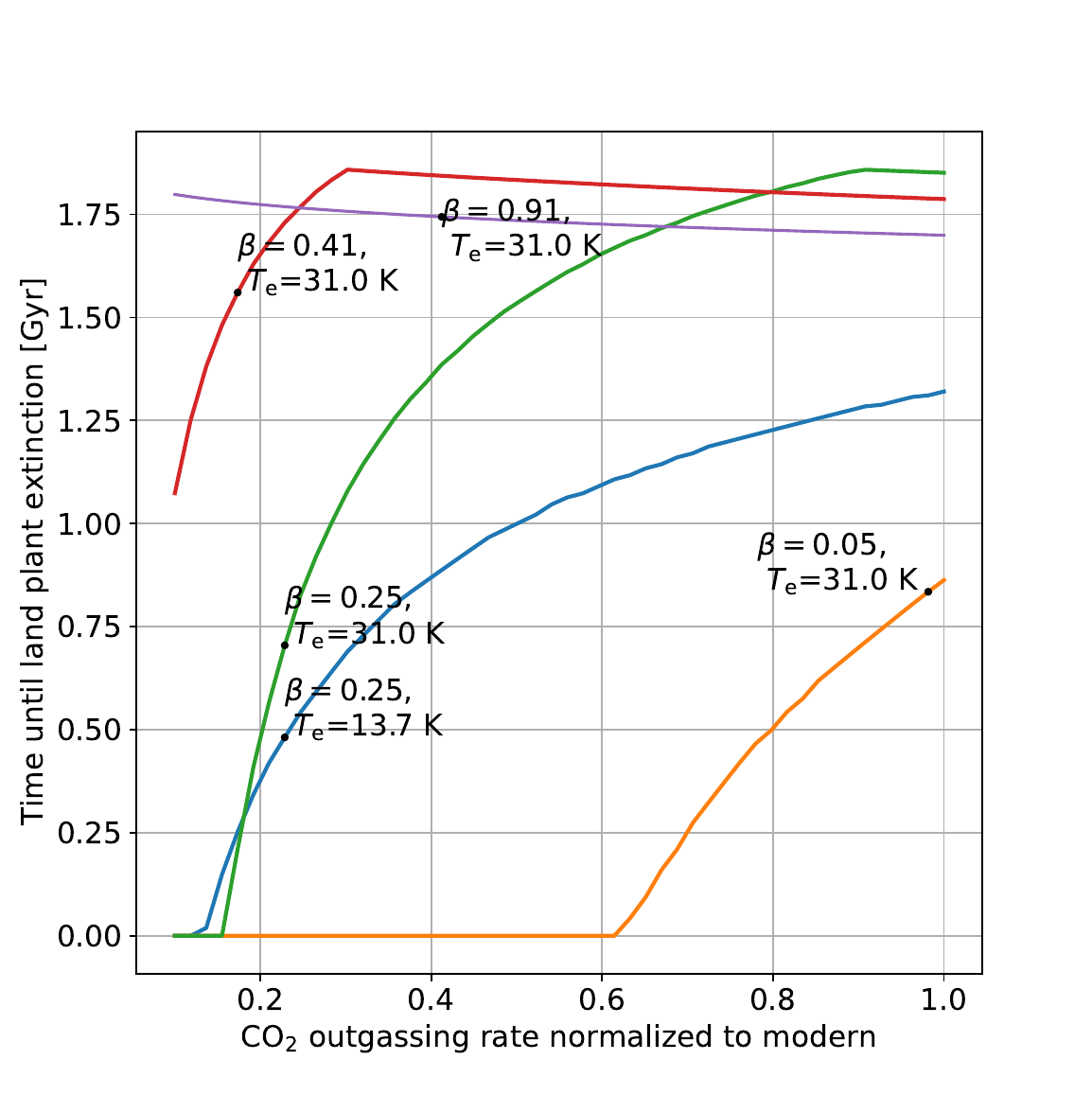}}
    \caption{\textbf{Time until land plant extinction [Gyr] vs. CO$_2$ outgassing rate relative to modern for varying weathering parameters.} The purple curve has $T_\mathrm{e}$=31.0 K and $\beta$=0.91, red has $T_\mathrm{e}$=31.0 K and $\beta$=0.41, orange has $T_\mathrm{e}$=31.0 K and $\beta$=0.05, green has $T_\mathrm{e}$=31.0 K and $\beta$=0.25, and blue has $T_\mathrm{e}$=13.7 K and $\beta$=0.25.}
    \label{fig:lifespan_outgassing}
\end{figure*}

The effect of outgassing on biosphere lifespan depends strongly, and independently, on $\beta$ and $T_\mathrm{e}$ (instead of just on their product). To demonstrate the variety in the response, we display biosphere lifespans as a function of outgassing rates down to 10\% of modern for a few combinations of weathering parameters (Fig.\,\ref{fig:lifespan_outgassing}). For simulations that die by overheating at modern outgassing (red curves in Fig.\,\ref{fig:lifespan_outgassing}), decreasing the outgassing rate initially causes an increasing lifespan by delaying the time of overheating. The delay lasts until the outgassing rate at which overheating coincides with the CO$_2$ starvation threshold is approached for a given set of parameters, producing a maximum lifespan of 1.86 Gyr. For very strong CO$_2$-dependence and weak temperature-dependence (e.g., with $\beta=0.91$, $T_\mathrm{e}=31.0$ K), even a reduction in outgassing of 90\% relative to modern does not shorten the lifespan at all. Changes in outgassing rate are, therefore, not likely to alter our conclusion that thermal stress is the more likely kill mechanism for the complex biosphere. In contrast, for outgassing rates and weathering parameters that produce CO$_2$ starvation (blue curves in Fig.\,\ref{fig:lifespan_outgassing}), any reduction in outgassing dramatically decreases the lifespan by allowing the minimum CO$_2$ for C$_4$ plants to be reached earlier.

\section{Discussion}
\subsection{Potential implications for life beyond Earth}\label{subsec:implications}
The future lifespan of the complex biosphere plays a key role in a body of literature that applies observational self-sampling assumptions to explain why technological intelligence appeared late in Earth's window of habitability and to estimate the number of critical, highly improbable steps in the evolution of intelligent life on Earth \citep{carter1983anthropic,hanson1998must,hanson1998great,flambaum_comment_2003,watson2004gaia,carter2008five,watson2008implications,cirkovic2009galactic,aldous2012great,bostrom2013anthropic,lingam2019role,snyder2021timing,hanson2021if,snyder2022catastrophe}. This statistical picture is frequently referred to as the ``Carter model'' \citep[e.g.][]{waltham2017star,snyder2021timing} after its originator Brandon Carter \citep{carter1983anthropic}. If the evolution of intelligent observers requires a series of $n$ critical, unlikely evolutionary transitions, each of which would typically be expected to take longer than the habitable lifetime of a planet, then if we condition on the success of all $n$ steps (with the $n$th step being the evolution of intelligence), the resulting probability density function for the emergence of intelligence as a function of time since the onset of habitability ($t$) can be written \citep[e.g.][]{watson2008implications}:
\begin{linenomath}
    \begin{equation}
        p_\mathrm{intel}(t) \approx \frac{nt^{n-1}}{\tau_\mathrm{hab}^n}
    \end{equation}
\end{linenomath}
where $\tau_\mathrm{hab}$ is the total duration of planetary habitability and other variables retain their previous definitions. From this, the expectation time for the evolution of an intelligent species can be derived:
\begin{linenomath}
    \begin{align}
    \tau_\mathrm{intel} &= \int_0^{\tau_\mathrm{hab}}tp_\mathrm{intel}(t)dt\\
    &= \frac{n}{n+1}\tau_\mathrm{hab}
    \end{align}
\end{linenomath}
which {implies} that a larger number $n$ of ``hard steps'' pushes the evolution of intelligence to later dates in the window of planetary habitability. {If we take the emergence of intelligence on Earth to coincide with the emergence of \textit{Homo sapiens}, then the time since the emergence of intelligence \citep[$10^5$-$10^6$ years;][]{galway2018did} is small relative to the gigayear timescales of Earth's biosphere's history and future, so} $\tau_\mathrm{hab}$ is simply the sum of $\tau_\mathrm{intel}$ and the future lifespan of the complex biosphere, which we will denote $\tau_\mathrm{future}${. With this simplification,} we can calculate $n$ as a function of timescales:
\begin{linenomath}
    \begin{align}
        \tau_\mathrm{intel} &= \tau_\mathrm{hab} - \tau_\mathrm{future}\\
        &= \frac{n+1}{n}\tau_\mathrm{intel} - \tau_\mathrm{future}\\
        n &= \frac{\tau_\mathrm{intel}}{\tau_\mathrm{future}}
    \end{align}
\end{linenomath}
meaning a longer future lifespan of the biosphere suggests proportionally fewer hard steps in the evolution of intelligent life \citep{carter1983anthropic,hanson1998must,carter2008five,watson2008implications,snyder2022catastrophe}.

Water was flowing on Earth's surface by 4.4 Ga \citep{valley2002cool, waltham2017star}, so we will take that as the time habitability began on Earth, with $\tau_\mathrm{intel} = 4.4$ Gyr following \citet{waltham2017star}, though the Late Heavy Bombardment may have prevented lasting habitability until 3.9 Ga \citep{pearce2018constraining}. With the estimate of $\tau_\mathrm{future}\approx1$ Gyr from \citet{Caldeira1992}, previous researchers have estimated $n\approx\frac{4.4}{1}\approx4-5$ \citep{carter2008five,watson2008implications,waltham2017star}. Alternatively, with our longest estimate for the future lifespan of $\tau_\mathrm{future}=1.86$ Gyr, the expected value for $n$ becomes $\frac{4.4}{1.86}\approx2.4$, suggesting two to three critical steps may be a better estimate. This would suggest that the emergence of intelligent life may be a less difficult (and consequently more common) process than some previous authors have argued \citep[e.g.][]{snyder2021timing,hanson2021if}, though since the hard steps can have arbitrarily small probabilities of occurring, intelligent life could still be extremely rare even with just a single hard step. 

The Carter model also allows for constraints to be placed on the timing of the hard steps that occurred  prior to the advent of intelligence. The probability density function and expected timing for step $m$ of an $n$-hard-step process can be written \citep{watson2008implications}:
\begin{linenomath}
\begin{align}
    p_{m/n}(t) &= \frac{n!}{(n-m)!(m-1)!}\frac{t^{m-1}(\tau_\mathrm{hab}-t)^{n-m}}{\tau_\mathrm{hab}^n}\label{eqn:pdf_m}\\
    \langle t_{m/n} \rangle &= \int_0^{\tau_\mathrm{hab}}tp_{m/n}(t)dt\\
     \langle t_{m/n} \rangle&= \frac{m}{n+1}\tau_\mathrm{hab}\label{eqn:exp_time_m}
\end{align}
\end{linenomath}
where $p_{m/n}(t)$ is the probability density function for step $m$ and $\langle t_{m/n} \rangle$ is the expected time between the onset of habitability and the occurrence of step $m$. Equation \ref{eqn:exp_time_m} makes it clear that the steps are distributed evenly across the habitable period, with each step separated on average by a duration of $\tau_\mathrm{hab}/(n+1)$, which is also the typical amount of time that separates step $n$ from the end of habitability, i.e. $\tau_\mathrm{hab}/(n+1)=\tau_\mathrm{future}$ and therefore
\begin{equation}
\langle t_{m/n}\rangle = m\tau_\mathrm{future}. 
\end{equation}

This equal spacing constraint means that the first hard step ought to take place approximately $\tau_\mathrm{future}$ years after the onset of habitability, which means the future lifespan of the biosphere imposes a timescale that can be used to evaluate whether the amount of time that passed between the onset of habitability and the origin of life on Earth ($\equiv\tau_\mathrm{OOL}$) is consistent with expected hard step timing. If $\tau_\mathrm{future}\approx\tau_\mathrm{OOL}$, then the origin of life is plausibly a hard step, meaning its typical timescale is  longer than the typical habitable period of Earth-like planets, suggesting life on other worlds will be rare. Alternatively, if $\tau_\mathrm{future}$ is significantly larger than $\tau_\mathrm{OOL}$, then this suggests that the origin of life is not a hard step (i.e. it typically happens on a shorter timescale than the lifetime of planetary habitability), which would imply that life might be fairly common in the universe. 

The timing of the onset of habitability and the timing of the origin of life on Earth are both somewhat poorly constrained \citep{pearce2018constraining}, resulting in estimates for $\tau_\mathrm{OOL}$ that could range from $\sim$0 Gyr to 0.8 Gyr, depending on which pieces of evidence are taken to confirm the presence of habitability and the presence of life. For example, if we take the the initiation of habitability on Earth to be 4.316 Ga based on an estimate for when water might have flowed on Earth after the Moon-forming impact under the assumption that Earth's magma ocean atmosphere cooled the surface inefficiently and took $\sim$0.1 Gyr to reach the condensation point of water at the surface \citep{pearce2018constraining} and we take the origin of life to be  4.28 Ga based on putative microfossils in early hydrothermal vent precipitates \citep{dodd2017evidence}, then $\tau_\mathrm{OOL}=4.316 - 4.28 = 0.036$ Gyr. Alternatively, if we assume that the Earth cooled efficiently after the Moon-forming impact, allowing water to flow earlier, then the start of habitability may have been as far back as 4.5 Ga, and if we take evidence for stromatolites 3.7 Ga as the earliest evidence for life, then $\tau_\mathrm{OOL}$ = 0.8 Gyr \citep{pearce2018constraining}.  

\begin{figure*}[htb!]
    \centering
    \makebox[\textwidth][c]{\includegraphics[width=270pt,keepaspectratio]{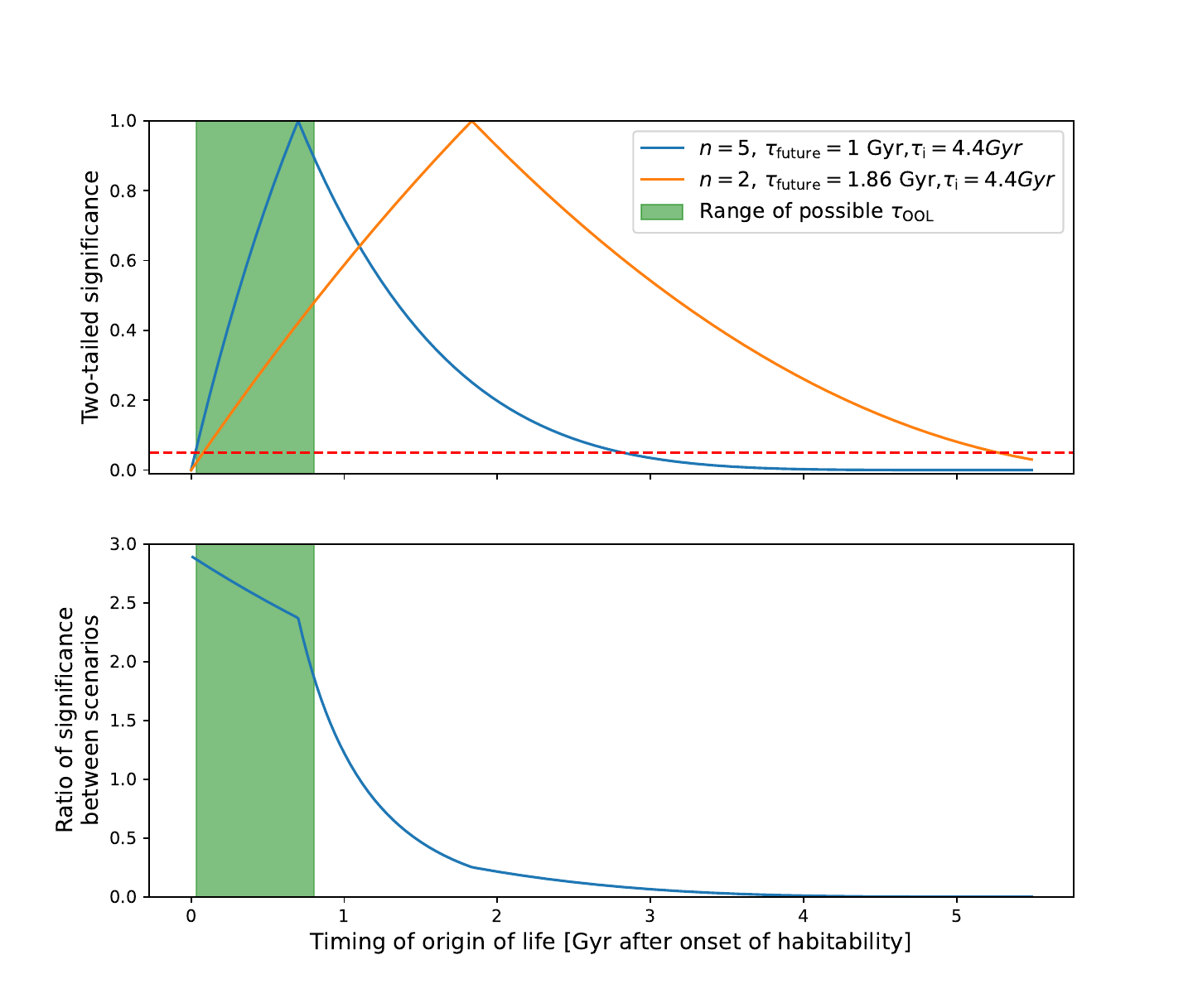}}
    \caption{\textbf{Significance and ratio of significance as a function of origin of life timing for two different values of the future lifespan of the biosphere and number of ``hard steps'' in the evolution of intelligence, assuming the origin of life was the first hard step.} (Top panel) The blue curve shows the significance as a function of timing for the origin of life as the first hard step in a $n=5$ hard step process in which $\tau_\mathrm{future}=1$ Gyr. The orange curve shows the significance as a function of timing for the origin of life as the first hard step in a $n=2$ hard step process in which $\tau_\mathrm{future}=1.86$ Gyr. The red dashed line denotes a 0.05 significance threshold. The shaded green area in the top and bottom panels shows a range of $\tau_\mathrm{OOL}$ permitted by evidence for the onset of habitability for Earth (defined by e.g. the first flowing water) and early evidence for the presence of life. The bottom panel shows the ratio of the significance values for the two scenarios described above, demonstrating that the likely $\tau_\mathrm{OOL}$ values are somewhat more surprising in the $n=5$ case, though not overwhelmingly so.}
    \label{fig:significance}
\end{figure*}

To quantify how surprising these timings would be if the origin of life was the first hard step in the evolution of intelligence, we can integrate the probability density function $p_{m/n}$ (Equation \ref{eqn:pdf_m}) to derive the cumulative probability $P_{m/n}$ of step $m=1$ as a function of time, then calculate the two-tailed significance with $2P_{m/n}$ if $P_{m/n}>0.5$ or $2(1-P_{m/n})$ if $P_{m/n}<0.5$ \citep{waltham2017star}. The smaller the calculated significance at a given time, the more surprising if would be for the origin of life to occur at that time, assuming it is a hard step. In the top panel of Figure \ref{fig:significance}, we plot the significance as a function of the timing of the origin of life with a model assuming $n=2$ hard steps with $\tau_\mathrm{future}=1.86$ Gyr (corresponding to our max calculation in this paper) and with a model assuming $n=5$ hard steps and $\tau_\mathrm{future}=1$ Gyr \citep[corresponding to previous calculations, e.g.][]{watson2008implications,carter2008five,waltham2017star}. The significance for the $n=5$ scenario ranges from 0.065 (or 6.5\%) for $\tau_\mathrm{OOL}=0.036$ Gyr up to a peak of 1 for $\tau_\mathrm{OOL}=0.7$ Gyr and back down to 0.897 for the maximum permitted $\tau_\mathrm{OOL}$ of 0.8 Gyr (blue line, top panel of Fig. \ref{fig:significance}). Using a common significance threshold of 0.05 \citep{waltham2017star}, none of these origin of life timings would be confidently ruled out in the $n=5$ scenario, meaning any of these timings would  be consistent with the origin of life being a hard step. For the $n=2$ scenario that arises from the longer future lifespan calculated in this study, the significance rises from 0.029 (2.9\%) at $\tau_\mathrm{OOL}=0.036$ Gyr to 0.48 at $\tau_\mathrm{OOL}=0.8$. The $n=2$ scenario reaches a significance of 0.05 at $\tau_\mathrm{OOL}=0.079$ Gyr, suggesting that if $\tau_\mathrm{OOL}$ were confirmed to lie between 0.036 and 0.079 Gyr, then the $n=2$ scenario would (very tentatively) suggest that the origin of life is not a hard step and thus may occur commonly on other (exo)planets. The bottom panel shows the ratio of the significance values of the two scenarios through time, demonstrating that the significance for the $n=5$, $\tau_\mathrm{future}=1$ Gyr scenario is 2-3 times larger than that of the $n=2$, $\tau_\mathrm{future}=1.86$ Gyr scenario across the given range of $\tau_\mathrm{OOL}$. These constraints are weak, but this suffices to show that a longer $\tau_\mathrm{future}$ may provide weak evidence that the origin of life is a common process, essentially by suggesting that intelligence arose less late in Earth's history than previously thought. Future investigations in the Carter model framework should account for the lengthened $\tau_\mathrm{future}$ suggested by this study. 

It may also be astrobiologically significant that a temperature upper bound for vascular plants of 338 K \citep{Redman2002,Marquez2007} may allow some forms of complex land life to survive to Earth's moist or runaway greenhouse transition \citep{Kasting1993,Leconte2013,Wolf2014,Wolf2015}. 3D climate models differ in their prediction of {climate evolution at high insolation and low CO$_2$}  \citep{Leconte2013,Wolf2014,Wolf2015,Popp2016,Wolf2018,Goldblatt2021, Yan2022}, with a range that includes our 0D climate model. {The moist greenhouse climate state (in which Earth's tropopause cold trap becomes less efficient, allowing the stratosphere to become moist and triggering geologically rapid loss of water to space) has been predicted to occur at surface temperatures ranging from 332 K \citep{Wolf2015} to 350 K \citep{kasting2015stratospheric}, meaning our threshold temperature for plant death approximately coincides with a possible mechanism for the loss of Earth's habitability. Similarly, the runaway greenhouse (in which Earth's ocean evaporates into the atmosphere at insolations above a critical threshold due to a limit on outgoing longwave radiation imposed by the interplay between water vapor's greenhouse effect and its exponential temperature sensitivity) has been estimated to occur at insolations between $S_\mathrm{eff}\approx1.1$ \citep{Leconte2013} (corresponding to 1.1 Gyr in the future) and $S_\mathrm{eff}>1.21$ \citep{Wolf2015} (corresponding to $>2.19$ Gyr in the future). Our predicting timings for land plant extinction also fall within this range of potential future runaway greenhouse timings.} Therefore, an important implication of our work is that the factors controlling Earth's transitions into exotic hot climate states {could be} a primary control on the lifespan of the complex biosphere, motivating further study of the moist and runaway greenhouse transitions with 3D models. Generalizing to exoplanets, this suggests that the inner edge of the ``complex life habitable zone'' may be coterminous with the inner edge of the classical circumstellar habitable zone, with relevance for where exoplanet astronomers might expect to find plant biosignatures like the ``vegetation red edge'' \citep{Seager:2005p2475}.

\subsection{Caveats}
{Our use of global-mean models prevents the representation of inherently 2- and 3-dimensional phenomena important for climate, silicate weathering, and plant growth. For example, 3D global climate model simulations suggest that spatially complex cloud feedbacks play a crucial role in determining the surface temperatures of hot, high-insolation climates like those expected in Earth's far future \citep{Wolf2015,Goldblatt2021,Yan2022}. The global-mean framework also obscures the fact that higher latitudes and altitudes tend to be cooler than the planet's mean temperature, potentially allowing land plants (and the organisms they sustain) to persist in shrinking polar and mountain refugia long past the point where Earth's global-mean temperature exceeds the temperature threshold for land plants \citep[``Swansong biospheres;''][]{omalley-james_swansong_2013}. Alterations in albedo and hydrological cycling driven by changes in the relative fractions of land covered by vegetation and desert as plant productivity falls may also influence Earth's far-future temperature and CO$_2$ trajectories \citep{Kleidon2000}. Additionally, the total global silicate weathering rate (and therefore the climate equilibrium of the carbon cycle) is sensitive to the spatial distribution of land on the planet \citep{baum2022sensitive}. A more computationally intensive modeling framework--e.g. a global climate model coupled to an interactive land model with dynamic vegetation--would be necessary to resolve effects like these and quantify their impact on the future lifespan of the biosphere.}

{Furthermore, even by the standards of global-mean parameterizations, the climate model we utilize in this study is quite simple and excludes several potentially relevant physical processes. For example, we assume a constant planetary Bond albedo of $a=0.29$, ignoring the potential for the loss of sea ice and accumulation of atmospheric water vapor to darken the planet under higher surface temperatures, positive feedbacks that which would accelerate warming and reduce the time required to reach either CO$_2$ starvation or the overheating threshold. We consider this justified for an idealized, global-mean climate model in the high-insolation regime studied here because, as noted in Section 
\ref{subsec:implications}, state-of-the-art 3D GCMs disagree in their predictions about Earth's climate response to increased insolation, particularly with respect to the impact of clouds on planetary albedo and longwave emission, which may have a destabilizing effect or a stabilizing effect on surface temperatures in response to the combination of increased insolation and decreased CO$_2$ \citep{Leconte2013,Wolf2014,Wolf2015,Popp2016,Wolf2018,Goldblatt2021, Yan2022}.} 

{Given the spread in predicted climates in state-of-the-art simulations, our model provides adequate estimates despite its various grave simplifications. For example, \citet{Leconte2013} used the the LMD GCM to predict a global-mean surface temperature of $\approx335$ K with an Earth-like 1 bar N$_2$ atmosphere and CO$_2$ of 375 ppmv at $S_\mathrm{eff}=1.1$, with any further increases to $S_\mathrm{eff}$ triggering a runaway greenhouse in their GCM, while \citet{Wolf2018} predicted a global-mean surface temperature of only 307 K for nearly the same conditions with 360 ppmv CO$_2$ using the Community Atmosphere Model version 4 (CAM4) GCM. For comparison, our global-mean model predicts a temperature of 311 K at modern insolation and 360 ppmv CO$_2$, intermediate between the LMD and CAM4 results, though closer to CAM4. Simulations using the Community Atmosphere Model version 3 (CAM3) GCM \citep{Wolf2014} produced global-mean surface temperatures of only 312.9 K with $S_\mathrm{eff}=$1.15, 500 ppmv CO$_2$, and 10 ppmv of CH$_4$, while our climate model predicts a significantly hotter 333 K temperature under those conditions, even without the extra greenhouse forcing provided by CH$_4$. Another set of simulations with CAM4 produced significantly warmer results than CAM3 \citep{Wolf2015}, with CAM4 predicting a temperature of 331.9 K at $S_\mathrm{eff}=$1.125 and 367 ppmv CO$_2$, while CAM3 predicted $\approx$305 K and our model predicts a temperature of 320.5 K for those conditions. Overall, our model's predictions fall within the wide range of those produced by more sophisticated GCMs. }

Most recent data (combining different subsets of paleoclimate proxy measurements, modern field observations, theoretical calculations, simulations, and laboratory experiments) suggest that silicate weathering is weakly temperature-dependent, with $T_\mathrm{e}\sim$30-40 K \citep{Maher2014,KrissansenTotton2017,Graham2020,Herbert2022,Brantley2023}, though one recent study examining the degree of chemical alteration of clays as a function of environmental variables suggested a strong dependence, with $T_\mathrm{e}\sim$12.1 K \citep{Deng2022}. Our work, therefore, cannot rule out with certainty CO$_2$ starvation as the complex biosphere kill mechanism. Our contribution should be read as a strong indication that the alternative mechanism of overheating is preferred, subject to further investigation of the global functioning of silicate weathering. 

We also note that we did not model photosynthesis by the third major photosynthetic pathway prevailing among land plants on Earth, crassulacean acid metabolism (CAM), which is used by roughly 6\% of land plant species \citep{Keeley2003}. Like C$_4$ plants, CAM plants concentrate CO$_2$ in their cells to enhance their rates of photosynthesis, but they use a different method. At night, they open their stomata to intake CO$_2$, storing the carbon as malic acid. During the day, stomata are closed to minimize water loss, and carbon is provided to Rubisco for light-driven fixation by decarboxylating the stored malic acid back into CO$_2$ \citep[e.g.][]{Keeley2003}. The strategy is adopted by many desert succulents to combat aridity \citep{Edwards2012}, and it would likely provide an increasingly significant advantage in the increasingly hot, low CO$_2$ climates expected in Earth's far future under a brightening Sun, since increased temperature will magnify the rate of evaporation from the plant, and reduced CO$_2$ will cause plants to open their stomata wider, also increasing water loss. {We excluded CAM plants from our analysis for two reasons: first, there is an apparent lack of simple temperature- and CO$_2$-dependent biochemical CAM productivity models analogous to those we used for C$_3$ and C$_4$ plants; and second, the particular adaptation of CAM plants to arid conditions would have necessitated an explicit model of water cycling, which is even less amenable to global-mean modeling than the existing components of our modeling framework. A 3D GCM coupled to a landscape model with dynamic vegetation would provide a more appropriate framework to test the impact of CAM photosynthesis on the future of land plants and serve as a natural extension of this work.}

Finally, environmental tolerance thresholds of land plants may only represent an evolutionary optimum under current conditions, rather than hard physical limits.  C$_4$ photosynthesis, which plays a major role in the ability of the biosphere to survive at low CO$_2$ in the far future in our calculations, appeared geologically recently, likely sometime in the Oligocene epoch between 24 and 35 Mya, potentially in response to the onset of extremely low CO$_2$ levels that have characterized the Earth ever since \citep{Ehleringer1997,Sage2004,Christin2011}. Because C$_4$ photosynthesis can operate under much lower ambient CO$_2$ levels than C$_3$, this innovation dramatically expanded the range of atmospheric conditions that would permit land plants to continue functioning on Earth. Since then, the C$_4$ metabolism has independently emerged in 19 families of plants, with some families displaying multiple originations for a total of at least 62 independent evolutionary lineages \citep{Sage2011}. Given the massive, continuing evolutionary changes that land plants have undergone since their origin $\approx$0.5 Ga, further evolution is plausible over the Gyr timescales considered in this article. One potential evolutionary pathway would be to combine existing carbon concentration mechanisms to ease the physiological stresses imposed by the potentially low CO$_2$ of the future. In fact, one species of succulent, \textit{Portulaca oleracea}, already displays an integrated C$_4$/CAM metabolism, demonstrating the potential viability of stacking carbon concentration mechanisms \citep{MorenoVillena2022}. Adaptation of land plants to a changing environment could push their extinction to even later dates than predicted here, {though the proximity of our maximum biosphere lifetime to predictions for the timing of the onset of the moist or runaway greenhouse suggests that the loss of Earth's oceans may limit the potential for further evolution of land plants to delay their extinction.}

\section{Conclusion}
In this study, we applied a global-mean model of Earth's {climate and} carbon cycle, which accounts for the  enhancement of silicate weathering by C$_3$ and C$_4$ plants, to re-evaluate the lifespan of the complex terrestrial biosphere under a brightening Sun. We show that recent data indicating weakly temperature-dependent silicate weathering \citep{Maher2014,KrissansenTotton2017,Graham2020,Herbert2022,Brantley2023} lead to the prediction that biosphere death results from overheating, not CO$_2$ starvation. These findings suggest that the future lifespan of Earth's complex biosphere may be nearly twice as long as previously thought. Specifically, if silicate weathering is weakly temperature-dependent and/or strongly CO$_2$-dependent, progressive decreases in plant productivity can slow, halt, and even temporarily reverse the expected future decrease in CO$_2$ as insolation continues to increase. Although this compromises the ability of the silicate weathering feedback to slow the warming of the Earth induced by higher insolation, it can also delay or prevent CO$_2$ starvation of land plants, allowing the continued existence of a complex land biosphere until the surface temperature becomes too hot. In this regime, contrary to previous results, expected future decreases in CO$_2$ outgassing and increases in land area would result in longer lifespans for the biosphere by delaying the point when land plants overheat. Importantly, with a revised thermotolerance limit for vascular land plants of 338 K, these results imply that the biotic feedback on weathering may allow complex land life to persist up to the moist {or runaway} greenhouse transition {on Earth (and potentially Earth-like exoplanets). Finally, as discussed in Section \ref{subsec:implications}, a longer future lifespan for the complex biosphere may also provide weak statistical evidence that there were fewer ``hard steps'' in the evolution of intelligent life than previously estimated and that the origin of life was not one of those hard steps.}
%% Also note that the akcnowlodgment environment does not support long amounts of text. If you have a lot of people and institutions to acknowledge, do not use this command. Instead, create a new \section{Acknowledgments}.
\section*{Acknowledgments}
This research was supported by a University of Chicago International Institute of Research in Paris Faculty Grant. Research was also sponsored by the National Aeronautics and Space Administration (NASA) through a contract with Oak Ridge Associated Universities (ORAU). The views and conclusions contained in this document are those of the authors and should not be interpreted as representing the official policies, either expressed or implied, of the National Aeronautics and Space Administration (NASA) or the U.S. Government. The U.S. Government is authorized to reproduce and distribute reprints for Government purposes notwithstanding any copyright notation herein. We thank Edwin Kite, Preston Kemeny, and at least 7 anonymous reviewers for providing useful critiques that significantly improved the paper's content and presentation.


\vspace{5mm}

\software{Matplotlib \citep{hunter2007matplotlib}, Numpy \citep{harris2020array}, Scipy \citep{jones2001scipy}          }

\appendix

\bibliography{biblio,biblio2}{}

\begin{thebibliography}{}
\expandafter\ifx\csname natexlab\endcsname\relax\def\natexlab#1{#1}\fi
\providecommand{\url}[1]{\href{#1}{#1}}
\providecommand{\dodoi}[1]{doi:~\href{http://doi.org/#1}{\nolinkurl{#1}}}
\providecommand{\doeprint}[1]{\href{http://ascl.net/#1}{\nolinkurl{http://ascl.net/#1}}}
\providecommand{\doarXiv}[1]{\href{https://arxiv.org/abs/#1}{\nolinkurl{https://arxiv.org/abs/#1}}}

\bibitem[{Abbot(2016)}]{Abbot2016}
Abbot, D.~S. 2016, The Astrophysical Journal, 827, 117

\bibitem[{Abbot {et~al.}(2012)Abbot, Cowan, \& Ciesla}]{D.S.Abbot2012}
Abbot, D.~S., Cowan, N.~B., \& Ciesla, F.~J. 2012, Astrophysical Journal, 756, 178

\bibitem[{Aldous(2012)}]{aldous2012great}
Aldous, D.~J. 2012, Mathematical Scientist, 37, 55

\bibitem[{Bahcall {et~al.}(2001)Bahcall, Pinsonneault, \& Basu}]{Bahcall2001}
Bahcall, J.~N., Pinsonneault, M., \& Basu, S. 2001, The Astrophysical Journal, 555, 990

\bibitem[{Baum {et~al.}(2022)Baum, Fu, \& Bourguet}]{baum2022sensitive}
Baum, M., Fu, M., \& Bourguet, S. 2022, Geophysical Research Letters, e2022GL098843

\bibitem[{Bernacchi {et~al.}(2001)Bernacchi, Singsaas, Pimentel, Portis~Jr, \& Long}]{Bernacchi2001}
Bernacchi, C., Singsaas, E., Pimentel, C., Portis~Jr, A., \& Long, S.~P. 2001, Plant, Cell \& Environment, 24, 253

\bibitem[{Berner {et~al.}(2003)Berner, Berner, \& Moulton}]{Berner2003}
Berner, E., Berner, R., \& Moulton, K. 2003, Treatise on geochemistry, 5, 605

\bibitem[{Berner(1992)}]{Berner1992}
Berner, R.~A. 1992, Geochimica et Cosmochimica Acta, 56, 3225

\bibitem[{Bj{\"o}rkman {et~al.}(1972)Bj{\"o}rkman, Pearcy, Harrison, \& Mooney}]{Bjoerkman1972}
Bj{\"o}rkman, O., Pearcy, R.~W., Harrison, A.~T., \& Mooney, H. 1972, Science, 175, 786

\bibitem[{Blunier {et~al.}(2012)Blunier, Bender, Barnett, \& Von~Fischer}]{Blunier2012}
Blunier, T., Bender, M., Barnett, B., \& Von~Fischer, J. 2012, Climate of the Past, 8, 1509

\bibitem[{Bostrom(2013)}]{bostrom2013anthropic}
Bostrom, N. 2013, Anthropic bias: Observation selection effects in science and philosophy (Routledge)

\bibitem[{Brantley {et~al.}(2023)Brantley, Shaughnessy, Lebedeva, \& Balashov}]{Brantley2023}
Brantley, S., Shaughnessy, A., Lebedeva, M.~I., \& Balashov, V.~N. 2023, Science, 379, 382

\bibitem[{Brown \& Smith(1975)}]{Brown1975}
Brown, W.~V., \& Smith, B.~N. 1975, Bulletin of the Torrey Botanical Club, 10

\bibitem[{Caldeira \& Kasting(1992)}]{Caldeira1992}
Caldeira, K., \& Kasting, J.~F. 1992, Nature, 360, 721

\bibitem[{Carter(1983)}]{carter1983anthropic}
Carter, B. 1983, Philosophical Transactions of the Royal Society of London. Series A, Mathematical and Physical Sciences, 310, 347

\bibitem[{Carter(2008)}]{carter2008five}
---. 2008, International Journal of Astrobiology, 7, 177

\bibitem[{Chen {et~al.}(1994)Chen, Coughenour, Knapp, \& Owensby}]{Chen1994}
Chen, D.-X., Coughenour, M.~B., Knapp, A.~K., \& Owensby, C.~E. 1994, Ecological Modelling, 73, 63, \dodoi{10.1016/0304-3800(94)90098-1}

\bibitem[{Chen {et~al.}(1970)Chen, Brown, \& Black}]{Chen1970}
Chen, T., Brown, R., \& Black, C. 1970, Weed science, 18, 399

\bibitem[{Christin {et~al.}(2011)Christin, Osborne, Sage, Arakaki, \& Edwards}]{Christin2011}
Christin, P.-A., Osborne, C.~P., Sage, R.~F., Arakaki, M., \& Edwards, E.~J. 2011, Journal of experimental Botany, 62, 3171

\bibitem[{{\'C}irkovi{\'c} {et~al.}(2009){\'C}irkovi{\'c}, Vukoti{\'c}, \& Dragi{\'c}evi{\'c}}]{cirkovic2009galactic}
{\'C}irkovi{\'c}, M.~M., Vukoti{\'c}, B., \& Dragi{\'c}evi{\'c}, I. 2009, Astrobiology, 9, 491

\bibitem[{Clarke(2014)}]{Clarke2014}
Clarke, A. 2014, International Journal of Astrobiology, 13, 141

\bibitem[{Collatz {et~al.}(1992)Collatz, Ribas-Carbo, \& Berry}]{Collatz1992}
Collatz, G.~J., Ribas-Carbo, M., \& Berry, J.~A. 1992, Functional Plant Biology, 19, 519

\bibitem[{Cox {et~al.}(1998)Cox, Huntingford, \& Harding}]{Cox1998}
Cox, P.~M., Huntingford, C., \& Harding, R.~J. 1998, Journal of Hydrology, 212-213, 79, \dodoi{10.1016/S0022-1694(98)00203-0}

\bibitem[{Coy(2022)}]{Coy2022}
Coy, B.~P. 2022, PhD thesis, UNIVERSITY OF CHICAGO

\bibitem[{Dahl \& Arens(2020)}]{Dahl2020}
Dahl, T.~W., \& Arens, S.~K. 2020, Chemical Geology, 547, 119665

\bibitem[{de~Sousa~Mello \& Fria{\c{c}}a(2020)}]{SousaMello2020}
de~Sousa~Mello, F., \& Fria{\c{c}}a, A. C.~S. 2020, International Journal of Astrobiology, 19, 25

\bibitem[{Deng {et~al.}(2022)Deng, Yang, \& Guo}]{Deng2022}
Deng, K., Yang, S., \& Guo, Y. 2022, Nature communications, 13, 1

\bibitem[{Dodd {et~al.}(2017)Dodd, Papineau, Grenne, Slack, Rittner, Pirajno, O’Neil, \& Little}]{dodd2017evidence}
Dodd, M.~S., Papineau, D., Grenne, T., {et~al.} 2017, Nature, 543, 60

\bibitem[{Edwards \& Ogburn(2012)}]{Edwards2012}
Edwards, E.~J., \& Ogburn, R.~M. 2012, International journal of plant sciences, 173, 724

\bibitem[{Ehleringer {et~al.}(1997)Ehleringer, Cerling, \& Helliker}]{Ehleringer1997}
Ehleringer, J.~R., Cerling, T.~E., \& Helliker, B.~R. 1997, Oecologia, 112, 285

\bibitem[{Farquhar {et~al.}(1980)Farquhar, von Caemmerer, \& Berry}]{Farquhar1980}
Farquhar, G.~D., von Caemmerer, S.~v., \& Berry, J.~A. 1980, planta, 149, 78

\bibitem[{Flambaum(2003)}]{flambaum_comment_2003}
Flambaum, V.~V. 2003, Astrobiology, 3, 237, \dodoi{10.1089/153110703769016307}

\bibitem[{Franck {et~al.}(2000)Franck, Block, Bloh, Bounama, Schellnhuber, \& Svirezhev}]{Franck2000}
Franck, S., Block, A., Bloh, W.~V., {et~al.} 2000, Tellus B, 52, 94

\bibitem[{Franck {et~al.}(2006)Franck, Bounama, \& Von~Bloh}]{Franck2006}
Franck, S., Bounama, C., \& Von~Bloh, W. 2006, Biogeosciences, 3, 85

\bibitem[{Franck {et~al.}(1999)Franck, Kossacki, \& Bounama}]{Franck1999}
Franck, S., Kossacki, K., \& Bounama, C. 1999, Chemical Geology, 159, 305

\bibitem[{Franck {et~al.}(2002)Franck, Kossacki, Von~Bloth, \& Bounama}]{Franck2002}
Franck, S., Kossacki, K.~J., Von~Bloth, W., \& Bounama, C. 2002, Tellus B: Chemical and Physical Meteorology, 54, 325

\bibitem[{Fran{\c{c}}ois {et~al.}(1998)Fran{\c{c}}ois, Delire, Warnant, \& Munhoven}]{Francois1998}
Fran{\c{c}}ois, L.~M., Delire, C., Warnant, P., \& Munhoven, G. 1998, Global and planetary change, 16, 37

\bibitem[{Galway-Witham \& Stringer(2018)}]{galway2018did}
Galway-Witham, J., \& Stringer, C. 2018, Science, 360, 1296

\bibitem[{Godsey {et~al.}(2009)Godsey, Kirchner, \& Clow}]{Godsey2009}
Godsey, S.~E., Kirchner, J.~W., \& Clow, D.~W. 2009, Hydrological Processes, 23, 1844, \dodoi{10.1002/hyp.7315}

\bibitem[{Goldblatt {et~al.}(2021)Goldblatt, McDonald, \& McCusker}]{Goldblatt2021}
Goldblatt, C., McDonald, V.~L., \& McCusker, K.~E. 2021, Nature Geoscience, 1

\bibitem[{Graham \& Pierrehumbert(2020)}]{Graham2020}
Graham, R., \& Pierrehumbert, R. 2020, Astrophysical Journal, 896

\bibitem[{Hakim {et~al.}(2021)Hakim, Bower, Tian, Deitrick, Auclair-Desrotour, Kitzmann, Dorn, Mezger, \& Heng}]{Hakim2021}
Hakim, K., Bower, D.~J., Tian, M., {et~al.} 2021, The Planetary Science Journal, 2, 49

\bibitem[{Hanson(1998{\natexlab{a}})}]{hanson1998must}
Hanson, R. 1998{\natexlab{a}}, Unpublished manuscript, September, 23, 168

\bibitem[{Hanson(1998{\natexlab{b}})}]{hanson1998great}
---. 1998{\natexlab{b}}, preprint available at http://hanson. gmu. edu/greatfilter. html

\bibitem[{Hanson {et~al.}(2021)Hanson, Martin, McCarter, \& Paulson}]{hanson2021if}
Hanson, R., Martin, D., McCarter, C., \& Paulson, J. 2021, The Astrophysical Journal, 922, 182

\bibitem[{Haqq-Misra {et~al.}(2016)Haqq-Misra, Kopparapu, Batalha, Harman, \& Kasting}]{HaqqMisra2016}
Haqq-Misra, J., Kopparapu, R.~K., Batalha, N.~E., Harman, C.~E., \& Kasting, J.~F. 2016, The Astrophysical Journal, 827, 120

\bibitem[{Harris {et~al.}(2020)Harris, Millman, Van Der~Walt, Gommers, Virtanen, Cournapeau, Wieser, Taylor, Berg, Smith, {et~al.}}]{harris2020array}
Harris, C.~R., Millman, K.~J., Van Der~Walt, S.~J., {et~al.} 2020, Nature, 585, 357

\bibitem[{Herbert {et~al.}(2022)Herbert, Dalton, Liu, Salazar, Si, \& Wilson}]{Herbert2022}
Herbert, T.~D., Dalton, C.~A., Liu, Z., {et~al.} 2022, Science, 377, 116

\bibitem[{Hunter(2007)}]{hunter2007matplotlib}
Hunter, J.~D. 2007, Computing in science \& engineering, 9, 90

\bibitem[{Ibarra {et~al.}(2019)Ibarra, Rugenstein, Bachan, Baresch, Lau, Thomas, Lee, Boyce, \& Chamberlain}]{Ibarra2019}
Ibarra, D.~E., Rugenstein, J. K.~C., Bachan, A., {et~al.} 2019, American Journal of Science, 319, 1

\bibitem[{Jacobs(1994)}]{Jacobs1994}
Jacobs, C. M.~J. 1994, Direct impact of atmospheric CO 2 enrichment on regional transpiration (Wageningen University and Research)

\bibitem[{Jones {et~al.}(2001)Jones, Oliphant, Peterson, {et~al.}}]{jones2001scipy}
Jones, E., Oliphant, T., Peterson, P., {et~al.} 2001

\bibitem[{Judson(2017)}]{Judson2017}
Judson, O.~P. 2017, Nature ecology \& evolution, 1, 0138

\bibitem[{Kasting {et~al.}(2015)Kasting, Chen, \& Kopparapu}]{kasting2015stratospheric}
Kasting, J.~F., Chen, H., \& Kopparapu, R.~K. 2015, The Astrophysical Journal Letters, 813, L3

\bibitem[{Kasting {et~al.}(1993)Kasting, Whitmire, \& Reynolds}]{Kasting1993}
Kasting, J.~F., Whitmire, D.~P., \& Reynolds, R.~T. 1993, Icarus, 101, 108

\bibitem[{Keeley \& Rundel(2003)}]{Keeley2003}
Keeley, J.~E., \& Rundel, P.~W. 2003, International journal of plant sciences, 164, S55

\bibitem[{Kleidon {et~al.}(2000)Kleidon, Fraedrich, \& Heimann}]{Kleidon2000}
Kleidon, A., Fraedrich, K., \& Heimann, M. 2000, Climatic Change, 44, 471, \dodoi{10.1023/A:1005559518889}

\bibitem[{Krissansen-Totton \& Catling(2017)}]{KrissansenTotton2017}
Krissansen-Totton, J., \& Catling, D.~C. 2017, Nature communications, 8, 1

\bibitem[{Lange {et~al.}(1974)Lange, Schulze, Evenari, Kappen, \& Buschbom}]{Lange1974}
Lange, O., Schulze, E.~D., Evenari, M., Kappen, L., \& Buschbom, U. 1974, Oecologia, 17, 97

\bibitem[{Leconte {et~al.}(2013)Leconte, Forget, Charnay, Wordsworth, \& Pottier}]{Leconte2013}
Leconte, J., Forget, F., Charnay, B., Wordsworth, R., \& Pottier, A. 2013, Nature, 504, 268

\bibitem[{Lenton \& von Bloh(2001)}]{Lenton2001}
Lenton, T.~M., \& von Bloh, W. 2001, Geophysical research letters, 28, 1715

\bibitem[{Lin {et~al.}(2012)Lin, Medlyn, \& Ellsworth}]{Lin2012}
Lin, Y.-S., Medlyn, B.~E., \& Ellsworth, D.~S. 2012, Tree physiology, 32, 219

\bibitem[{Lingam \& Loeb(2019)}]{lingam2019role}
Lingam, M., \& Loeb, A. 2019, International Journal of Astrobiology, 18, 527

\bibitem[{Lovelock \& Whitfield(1982)}]{Lovelock1982}
Lovelock, J.~E., \& Whitfield, M. 1982, Nature, 296, 561

\bibitem[{Maher \& Chamberlain(2014)}]{Maher2014}
Maher, K., \& Chamberlain, C. 2014, Science, 343, 1502

\bibitem[{Manabe {et~al.}(2004)Manabe, Wetherald, Milly, Delworth, \& Stouffer}]{Manabe2004}
Manabe, S., Wetherald, R.~T., Milly, P. C.~D., Delworth, T.~L., \& Stouffer, R.~J. 2004, Climatic Change, 64, 59, \dodoi{10.1023/B:CLIM.0000024674.37725.ca}

\bibitem[{M{\'a}rquez {et~al.}(2007)M{\'a}rquez, Redman, Rodriguez, \& Roossinck}]{Marquez2007}
M{\'a}rquez, L.~M., Redman, R.~S., Rodriguez, R.~J., \& Roossinck, M.~J. 2007, science, 315, 513

\bibitem[{Medlyn {et~al.}(2011)Medlyn, Duursma, Eamus, Ellsworth, Prentice, Barton, Crous, De~Angelis, Freeman, \& Wingate}]{Medlyn2011}
Medlyn, B.~E., Duursma, R.~A., Eamus, D., {et~al.} 2011, Global Change Biology, 17, 2134

\bibitem[{Medlyn {et~al.}(2012)Medlyn, Duursma, Eamus, Ellsworth, Colin~Prentice, Barton, Crous, Angelis, Freeman, \& Wingate}]{Medlyn2012}
---. 2012, Global Change Biology, 18, 3476

\bibitem[{Mello \& Friaça(2023)}]{Mello2023}
Mello, F. d.~S., \& Friaça, A. C.~S. 2023, International Journal of Astrobiology, 1, \dodoi{10.1017/S1473550423000083}

\bibitem[{Miller {et~al.}(2013)Miller, McGuirl, \& Carvey}]{Miller2013}
Miller, S.~R., McGuirl, M.~A., \& Carvey, D. 2013, Molecular biology and evolution, 30, 752

\bibitem[{Moreno-Villena {et~al.}(2022)Moreno-Villena, Zhou, Gilman, Tausta, Cheung, \& Edwards}]{MorenoVillena2022}
Moreno-Villena, J.~J., Zhou, H., Gilman, I.~S., {et~al.} 2022, Science Advances, 8, eabn2349

\bibitem[{Moulton \& Berner(1998)}]{Moulton1998}
Moulton, K.~L., \& Berner, R.~A. 1998, Geology, 26, 895

\bibitem[{Moulton {et~al.}(2000)Moulton, West, \& Berner}]{Moulton2000}
Moulton, K.~L., West, J., \& Berner, R.~A. 2000, American Journal of Science, 300, 539

\bibitem[{Nobel(2020)}]{Nobel2020}
Nobel, P.~S. 2020, in Physicochemical and {Environmental} {Plant} {Physiology} (Elsevier), 409--488, \dodoi{10.1016/B978-0-12-819146-0.00008-0}

\bibitem[{O'Malley-James {et~al.}(2013)O'Malley-James, Greaves, Raven, \& Cockell}]{omalley-james_swansong_2013}
O'Malley-James, J.~T., Greaves, J.~S., Raven, J.~A., \& Cockell, C.~S. 2013, International Journal of Astrobiology, 12, 99, \dodoi{10.1017/S147355041200047X}

\bibitem[{Ozaki \& Reinhard(2021)}]{Ozaki2021}
Ozaki, K., \& Reinhard, C.~T. 2021, Nature Geoscience, 14, 138

\bibitem[{Palandri \& Kharaka(2004)}]{Palandri2004}
Palandri, J.~L., \& Kharaka, Y.~K. 2004, A compilation of rate parameters of water-mineral interaction kinetics for application to geochemical modeling, Tech. rep.

\bibitem[{Pearce {et~al.}(2018)Pearce, Tupper, Pudritz, \& Higgs}]{pearce2018constraining}
Pearce, B.~K., Tupper, A.~S., Pudritz, R.~E., \& Higgs, P.~G. 2018, Astrobiology, 18, 343

\bibitem[{Pennisi(2003)}]{Pennisi2003}
Pennisi, E. 2003, Fungi shield new host plants from heat and drought,  American Association for the Advancement of Science

\bibitem[{Perkins(2021)}]{Perkins2021}
Perkins, J. 2021, Field Studies in Ecology, 3

\bibitem[{Popp {et~al.}(2016)Popp, Schmidt, \& Marotzke}]{Popp2016}
Popp, M., Schmidt, H., \& Marotzke, J. 2016, Nature Communications, 7, 10627

\bibitem[{Prado {et~al.}(2023)Prado, Xue, Johnson, Field, Stata, Hawkins, Hsia, Liu, Cheng, \& Rhee}]{Prado2023}
Prado, K., Xue, B., Johnson, J.~E., {et~al.} 2023, bioRxiv, 2023

\bibitem[{Raven {et~al.}(2008)Raven, Cockell, \& De~La~Rocha}]{Raven2008}
Raven, J.~A., Cockell, C.~S., \& De~La~Rocha, C.~L. 2008, Philosophical Transactions of the Royal Society B: Biological Sciences, 363, 2641, \dodoi{10.1098/rstb.2008.0020}

\bibitem[{Redman {et~al.}(2002)Redman, Sheehan, Stout, Rodriguez, \& Henson}]{Redman2002}
Redman, R.~S., Sheehan, K.~B., Stout, R.~G., Rodriguez, R.~J., \& Henson, J.~M. 2002, Science, 298, 1581

\bibitem[{Rogers \& Humphries(2000)}]{Rogers2000}
Rogers, A., \& Humphries, S.~W. 2000, Global Change Biology, 6, 1005, \dodoi{10.1046/j.1365-2486.2000.00375.x}

\bibitem[{Rushby {et~al.}(2018)Rushby, Johnson, Mills, Watson, \& Claire}]{Rushby2018}
Rushby, A.~J., Johnson, M., Mills, B.~J., Watson, A.~J., \& Claire, M.~W. 2018, Astrobiology, 18, 469

\bibitem[{Sage(2004)}]{Sage2004}
Sage, R.~F. 2004, New Phytologist, 161, 341, \dodoi{10.1111/j.1469-8137.2004.00974.x}

\bibitem[{Sage {et~al.}(2011)Sage, Christin, \& Edwards}]{Sage2011}
Sage, R.~F., Christin, P.-A., \& Edwards, E.~J. 2011, Journal of experimental botany, 62, 3155

\bibitem[{Salvucci \& Crafts-Brandner(2004)}]{Salvucci2004}
Salvucci, M.~E., \& Crafts-Brandner, S.~J. 2004, Plant physiology, 134, 1460

\bibitem[{Salvucci {et~al.}(2001)Salvucci, Osteryoung, Crafts-Brandner, \& Vierling}]{Salvucci2001}
Salvucci, M.~E., Osteryoung, K.~W., Crafts-Brandner, S.~J., \& Vierling, E. 2001, Plant physiology, 127, 1053

\bibitem[{Scafaro {et~al.}(2019)Scafaro, Bautsoens, Den~Boer, Van~Rie, \& Gall{\'e}}]{Scafaro2019}
Scafaro, A.~P., Bautsoens, N., Den~Boer, B., Van~Rie, J., \& Gall{\'e}, A. 2019, Plant Physiology, 181, 43

\bibitem[{Scafaro {et~al.}(2023)Scafaro, Posch, Evans, Farquhar, \& Atkin}]{Scafaro2023}
Scafaro, A.~P., Posch, B.~C., Evans, J.~R., Farquhar, G.~D., \& Atkin, O.~K. 2023, Nature Communications, 14, 2820

\bibitem[{Schwartzman(2017)}]{Schwartzman2017}
Schwartzman, D.~W. 2017, AIMS Geosciences, 3, 216, \dodoi{10.3934/geosci.2017.2.216}

\bibitem[{Seager {et~al.}(2005)Seager, Turner, Schafer, \& Ford}]{Seager:2005p2475}
Seager, S., Turner, E., Schafer, J., \& Ford, E. 2005, Astrobiology, 5, 372

\bibitem[{Shivhare \& Mueller-Cajar(2017)}]{Shivhare2017}
Shivhare, D., \& Mueller-Cajar, O. 2017, Plant Physiology, 174, 1505

\bibitem[{Snyder-Beattie \& Bonsall(2022)}]{snyder2022catastrophe}
Snyder-Beattie, A.~E., \& Bonsall, M.~B. 2022, Proceedings of the Royal Society B, 289, 20212711

\bibitem[{Snyder-Beattie {et~al.}(2021)Snyder-Beattie, Sandberg, Drexler, \& Bonsall}]{snyder2021timing}
Snyder-Beattie, A.~E., Sandberg, A., Drexler, K.~E., \& Bonsall, M.~B. 2021, Astrobiology, 21, 265

\bibitem[{Stephens {et~al.}(2015)Stephens, O'Brien, Webster, Pilewski, Kato, \& Li}]{Stephens2015}
Stephens, G.~L., O'Brien, D., Webster, P.~J., {et~al.} 2015, Reviews of geophysics, 53, 141

\bibitem[{Tansey \& Brock(1972)}]{Tansey1972}
Tansey, M.~R., \& Brock, T.~D. 1972, Proceedings of the National Academy of Sciences, 69, 2426

\bibitem[{Taylor {et~al.}(2011)Taylor, Banwart, Leake, \& Beerling}]{Taylor2011}
Taylor, L., Banwart, S., Leake, J., \& Beerling, D.~J. 2011, American Journal of Science, 311, 369, \dodoi{10.2475/05.2011.01}

\bibitem[{Taylor {et~al.}(2009)Taylor, Leake, Quirk, Hardy, Banwart, \& Beerling}]{Taylor2009}
Taylor, L., Leake, J., Quirk, J., {et~al.} 2009, Geobiology, 7, 171

\bibitem[{Valley {et~al.}(2002)Valley, Peck, King, \& Wilde}]{valley2002cool}
Valley, J.~W., Peck, W.~H., King, E.~M., \& Wilde, S.~A. 2002, Geology, 30, 351

\bibitem[{Volk(1987)}]{Volk1987}
Volk, T. 1987, in America Journal of Science, 763--779

\bibitem[{Von~Bloh {et~al.}(2003)Von~Bloh, Franck, Bounama, \& Schellnhuber}]{VonBloh2003}
Von~Bloh, W., Franck, S., Bounama, C., \& Schellnhuber, H.-J. 2003, Geomicrobiology Journal, 20, 501

\bibitem[{Walker {et~al.}(1981)Walker, Hays, \& Kasting}]{Walker1981}
Walker, J.~C.~G., Hays, P.~B., \& Kasting, J.~F. 1981, Journal of Geophysical Research, 86, 9776

\bibitem[{Waltham(2017)}]{waltham2017star}
Waltham, D. 2017, Astrobiology, 17, 61

\bibitem[{Watson(2004)}]{watson2004gaia}
Watson, A. 2004, in Scientists debate Gaia: the next century (MIT Press), 201--208

\bibitem[{Watson(2008)}]{watson2008implications}
Watson, A.~J. 2008, Astrobiology, 8, 175

\bibitem[{Winnick \& Maher(2018)}]{Winnick2018}
Winnick, M.~J., \& Maher, K. 2018, Earth and Planetary Science Letters, 485, 111

\bibitem[{Wolf {et~al.}(2018)Wolf, Haqq-Misra, \& Toon}]{Wolf2018}
Wolf, E., Haqq-Misra, J., \& Toon, O. 2018, Journal of Geophysical Research: Atmospheres, 123, 11

\bibitem[{Wolf \& Toon(2014)}]{Wolf2014}
Wolf, E., \& Toon, O. 2014, Geophysical Research Letters, 41, 167

\bibitem[{Wolf \& Toon(2015)}]{Wolf2015}
---. 2015, Journal of Geophysical Research, 120, 5775

\bibitem[{Wong {et~al.}(1979)Wong, Cowan, \& Farquhar}]{Wong1979}
Wong, S., Cowan, I., \& Farquhar, G. 1979, Nature, 282, 424

\bibitem[{Yan {et~al.}(2022)Yan, Yang, Zhang, \& Huang}]{Yan2022}
Yan, M., Yang, J., Zhang, Y., \& Huang, H. 2022, Geophysical Research Letters, 49, e2022GL100152, \dodoi{10.1029/2022GL100152}

\bibitem[{Yang {et~al.}(2022)Yang, Brandon, Landais, Duchamp-Alphonse, Blunier, Pri{\'e}, \& Extier}]{Yang2022}
Yang, J.-W., Brandon, M., Landais, A., {et~al.} 2022, Science, 375, 1145

\end{thebibliography}
\bibliographystyle{aasjournal}



\end{document}